\documentclass[aps,prd,twocolumn,groupedaddress,nofootinbib,longbibliography]{revtex4-1}

\usepackage{amssymb}
\usepackage{amsmath}
\usepackage[linktocpage=true]{hyperref}
\usepackage{graphicx}

\begin{document}

\title{Static axisymmetric rings in general relativity: How diverse they are}

\author{O. Semer\'ak}
\email[]{oldrich.semerak@mff.cuni.cz}

\affiliation{Institute of Theoretical Physics, Faculty of Mathematics and Physics, Charles University, Prague, Czech Republic}

\date{\today}

\begin{abstract}
Three static and axially symmetric (Weyl-type) ring singularities -- the Majumdar-Papapetrou--type (extremally charged) ring, the Bach-Weyl ring and the Appell ring -- are studied in general relativity in order to show how remarkably the geometries in their vicinity differ from each other. This is demonstrated on basic measures of the rings and on invariant characteristics given by the metric and by its first and second derivatives (lapse, gravitational acceleration and curvature), and also on geodesic motion. The results are also compared against the Kerr space-time which possesses a ring singularity too. The Kerr solution is only stationary, not static, but in spite of the consequent complication by dragging, its ring appears to be simpler than the static rings. We show that this mainly applies to the Bach-Weyl ring, although this straightforward counter-part of the Newtonian homogeneous circular ring is by default being taken as the simplest ring solution, and although the other two static ring sources may seem more ``artificial''. The weird, directional deformation around the Bach-Weyl ring probably indicates that a more adequate coordinate representation and interpretation of this source should exists.
\end{abstract}


\maketitle

\section{Introduction}

``Apparently, the rich structure of the Weyl-type solutions is far from completely explored.'' is the last sentence of paper \cite{Shoom-15b} and we beg to borrow it as the first one.\footnote
{It is especially inviting to do so in the eve of the 100 years of Weyl solutions \cite{Weyl-17}.}
The quoted paper deals with stationary situation, but the statement even holds for a static case which will be considered here. More specifically, we will focus on static, axially symmetric and asymptotically flat gravitational fields generated by thin circular rings. The solution of Einstein's equations may then seem to simply lead to the aged Bach-Weyl solution which just adds, to the well-known Newtonian potential given by an elliptic integral, ``the second'' metric function obtained by a line integral given by the potential gradient. However, (i) exactly this second function (involved in the meridional-plane geometry) makes the field very different from the Newtonian one, mainly in the vicinity of the ring, and (ii) there exist various other Weyl-class solutions with ring-singularity source which differ from the Bach-Weyl solution considerably. In the present paper, we demonstrate the differences on several basic geometrical properties, thus reminding that at least some of the sources are not ``simple rings'' (this mainly applies to the Bach-Weyl ring).

In the following paper, we plan to check whether the rings just represent naked singularities, or whether apparent horizons can appear around them under some conditions (which would indicate that the solution in fact cannot stay static, namely that the ring has to shrink to a point if ``released'' from the time-symmetry configuration). A number of results have already been obtained in this respect, but let us mention them only then.

The first solution we choose, besides that due to Bach \& Weyl, is the ring obtained as a continuous limit of a circular-ring distribution of the Majumdar-Papapetrou extreme black holes. This solution is not vacuum of course (it contains an electromagnetic field) and may seem rather artificial, but its properties will actually be found more satisfactory than those of the Bach-Weyl solution. Another ring to compare will be the Appell solution, originally known from electrodynamics (already in the 19th century) and much more recently also introduced into general relativity. In contrast to the previous two, it extends over {\em two} asymptotically flat domains, connected via a circle spanned by the ring. A very similar double-sheeted topology is know from the Kerr solution which is generated by a singular ring as well, so we also subject the latter to this comparison, although it is {\em rotating}, thus only stationary, not static.

Below, in section \ref{metric,coords}, we first write the Weyl-type metric in several useful coordinates. Then, in section \ref{ring-sources}, the basic data are listed about the rings we will compare. Several simple properties of the rings are calculated in section \ref{simple-properties} indicating how the spatial geometry behaves in their vicinity. In section \ref{geod-acceleration}, we compare the geometries on two simple types of geodesics, and then by computing the gravitational acceleration, given by gradient of the lapse function. Curvature generated by the rings (the Gauss curvature of several privileged two-surfaces and the space-time Kretschmann scalar) is examined in section \ref{curvature}. The properties are illustrated by figures, in particular, the geometry is visualized on contour plots (with graduated shading) of the scalars obtained from the metric (lapse) and from its derivatives (acceleration and curvature). Remarks not included directly at respective places are added in section \ref{concluding}.

The main aim of this study is to demonstrate, on several examples, that already within very simple -- static, axially (and reflectionally) symmetric and asymptotically flat -- class, the exact thin-ring solutions of general relativity differ from each other considerably. It is clear that being infinitesimally thin, such rings represent (naked) curvature singularities, so cannot be considered (astro)physically realistic in their closest vicinity. The latter is in fact already true within Newtonian theory, because thin (volume-less) sources involve infinite density, hence diverging derivatives of the gravitational field. However, some of the relativistic rings add more serious pathology to this factually natural behaviour, namely their properties are strongly dependent on direction from which they are approached (they are far from being locally cylindrical). We will call such sources less ``reasonable'' or ``satisfactory'', which will precisely mean that they are not {\it simple} in the sense of Israel \cite{Israel-77}, or, in the terms of \cite{Liang-73}, they are not ``normal-dominated'' singularities. It should be add that such a behaviour need not disqualify the source completely, since it may be tied to a given particular type of representation (in our case the one based on the Weyl coordinates). Indeed, it has been proposed how some of such pathologies may be transformed out, as further commented on in concluding remarks.

Below, we use geometrized units in which $c=1$, $G=1$, index-posed comma/semicolon indicates partial/covariant derivative and usual summation rule is employed. Signature of the space-time metric $g_{\mu\nu}$ is ($-$+++), Riemann tensor is defined according to $V_{\nu;\kappa\lambda}-V_{\nu;\lambda\kappa}={R^\mu}_{\nu\kappa\lambda}V_\mu$
and Ricci tensor by $R_{\nu\lambda}={R^\kappa}_{\nu\kappa\lambda}$.

\section{Weyl metric, useful coordinates}
\label{metric,coords}

Assuming the configuration to be static and axially symmetric, we consider as basic the Weyl-type cylindrical coordinates $(t,\rho,\phi,z)$ of which $t$ and $\phi$ are parameters of the two Killing symmetries, $z\!=\!0$ is the ring plane and $\rho\!=\!0$ represents its symmetry axis. The metric can be written
\begin{equation}  \label{metric-Weyl}
  {\rm d}s^2=-N^2{\rm d}t^2+N^{-2}\left[\rho^2{\rm d}\phi^2+e^{2\lambda}({\rm d}\rho^2+{\rm d}z^2)\right],
\end{equation}
where the lapse $N$ is often being expressed as $N\!\equiv\!e^\nu$ in terms of a gravitational potential $\nu$ which satisfies Laplace or Poisson equation. The second metric function $\lambda$ is determined by line integral of an expression given by gradient of $\nu$; it is zero on the symmetry axis (at least if no source is there).

Aiming to study the fields generated by circular thin rings, we will naturally set the coordinates so that the rings lie in the ``equatorial plane'' $z\!=\!0$, on some radius $\rho\!=\!a$.\footnote
{The ring radius $a$ is assumed to be positive and finite in general. The limit $a\!\rightarrow\!0^+$ is an issue on its own which we do not tackle here. The limit of the ring solutions themselves is clear -- we will see that for the Majumdar-Papapetrou ring it is the extreme Reissner-Nordstr\"om horizon, for the Bach-Weyl and Appell rings it is the Curzon ``directional-particle'' source, and for Kerr it is of course Schwarzschild -- but much less clear is the limit of various formulas given below: apparently some of them even work in this limit, whereas some do not. We occasionally add a remark in this respect, but often it is so that a formula obtained for $0\!<\!a\!<\!\infty$ does {\em not} go over, in the $a\!\rightarrow\!0^+$ limit, to the result obtained ``from the beginning'' for the corresponding (above mentioned) limit space-time.}
Regarding this setting, it is often suitable to transform to toroidal coordinates with ``focus'' at the ring,
\begin{equation}  \label{toroidal-coords}
  \rho=\frac{a\,\sinh\zeta}{\cosh\zeta-\cos\psi} \,, \quad
  z=\frac{a\,\sin\psi}{\cosh\zeta-\cos\psi} \,,
\end{equation}
where $\zeta$ is the new radius ($0\!\leq\!\zeta\!<\!\infty$) and $\psi$ is the new latitudinal coordinate ($0\!\leq\!\psi\!<\!2\pi$).
In such coordinates, the metric takes the form
\begin{equation}  \label{metric-toroidal}
  {\rm d}s^2=
  -N^2{\rm d}t^2
  +\frac{a^2}{N^2}\,
   \frac{\sinh^2\zeta\,{\rm d}\phi^2+e^{2\lambda}({\rm d}\zeta^2+{\rm d}\psi^2)}
        {(\cosh\zeta-\cos\psi)^2} \;.
\end{equation}
Toroidal coordinates are mainly appropriate if one needs to approach the ring from a generic direction.

The third useful coordinates are the ellipsoidal (oblate spheroidal) ones, $R$ and $\vartheta$, related to the Weyl coordinates by
\begin{equation}  \label{oblate}
  \rho=\sqrt{R^2+a^2}\,\sin\vartheta, \qquad
  z=R\cos\vartheta.
\end{equation}
The metric reads in them
\begin{align}
  {\rm d}s^2
  = & -N^2\,{\rm d}t^2
      +\frac{(R^2+a^2)\sin^2\vartheta}{N^2}\,{\rm d}\phi^2+ \nonumber \\
    & +\frac{\Sigma\,e^{2\lambda}}{N^2}
       \left(\frac{{\rm d}R^2}{R^2+a^2}+{\rm d}\vartheta^2\right),
  \label{metric-oblate}
\end{align}
where
\begin{align}
  l_{1,2}&:=\sqrt{(\rho\mp a)^2+z^2}
           =\sqrt{\frac{2a^2 e^{\mp\zeta}}{\cosh\zeta-\cos\psi}} \nonumber \\
         & \;=\sqrt{R^2+a^2}\mp a\sin\vartheta \,, \nonumber \\
  \Sigma &:=l_1 l_2
           =\sqrt{(\rho^2-a^2+z^2)^2+4a^2 z^2}  \nonumber \\
         & \;=\frac{2a^2}{\cosh\zeta-\cos\psi}
           =R^2+a^2\cos^2\vartheta \,,  \label{Sigma}
\end{align}
and
\begin{equation}
  \sqrt{R^2+a^2}=\frac{l_2+l_1}{2} \;, \qquad
  a\,\sin\vartheta=\frac{l_2-l_1}{2}
\end{equation}
is the inverse transformation.

Note that at the Kerr solution exactly the same transformation as (\ref{oblate}) will be used, but in that case $(\rho,z)$ will stand for the Kerr-Schild cylindrical coordinates (rather than for the Weyl ones) and the oblate coordinates will be represented by Boyer-Lindquist ones. The Boyer-Lindquist coordinates are oblate with respect to the Kerr-Schild coordinates, but {\em prolate} with respect to the Weyl ones, yet we still keep the notation $(\rho,z)$ in both cases (because for static rings only Weyl coordinates will be employed, while for Kerr only the Kerr-Schild ones). One could similarly keep the same notation for the above oblate coordinates as for the Boyer-Lindquist ones, because {\em in this paper} they will play an analogous role, yet we have decided to distinguish them by using $R$, $\vartheta$, while the Boyer-Lindquist ones will be denoted by $r$, $\theta$ as usual. The reason is that the Boyer-Lindquist coordinates are ``Schwarzschild-type'' (they reduce to the Schwarzschild coordinates if the radius of the Kerr ring shrinks to zero), whereas the oblate coordinates introduced by (\ref{oblate}) are different: with respect to the Weyl coordinates, the Schwarzschild coordinates are {\em prolate}, being related by
$\rho=\sqrt{r(r\!-\!2M)}\,\sin\theta$, $z=(r\!-\!M)\cos\theta$.

\section{Ring sources we will compare}
\label{ring-sources}

Below we list the basic data about the rings we will compare -- the one obtained as a continuous circular distribution 
of Majumdar-Papapetrou--type (extremally charged) particles, the Bach-Weyl ring, the Appell ring and the Kerr ring. The Kerr ring is stationary but not static, and it generates a Killing horizon if its radius is not bigger than its mass, yet it is natural to include it, because it is a prominent ring singularity and it has some common features with the Appell ring.

\subsection{Majumdar-Papapetrou (MP) ring}

It is known (e.g. \cite{ChruscielCH-12}) that the Majumdar-Papapetrou solutions provide the only known singularity-free stationary (in fact even static) electrovacuum space-times with more than one black hole. 
Their metric has $\lambda\!=\!0$ everywhere and is usually presented as
\begin{equation}
  {\rm d}s^2=-N^2{\rm d}t^2+N^{-2}({\rm d}x^2+{\rm d}y^2+{\rm d}z^2)
\end{equation}
in Cartesian-type coordinates $(x,y,z)$ related to the Weyl coordinates by
\[x=\rho\,\cos\phi, \quad y=\rho\,\sin\phi \,.\]
The lapse function $N$ is given by
\[\frac{1}{N}=1+\sum_{j=1}^{n}\frac{M_j}{|\vec{r}-\vec{r}_j|} \;,\]
$n$ being the number of black holes and $M_j$ and $\vec{r}_j\equiv(x_j,y_j,z_j)$ denoting their masses and positions (specifically, the positions of their horizons which are represented as points in the above coordinates). The black holes are just in stationary equilibrium thanks to their extremal electric charges ($|Q_i|\!=\!M_i$ in geometrized units), which implies that the electromagnetic field is given by potential $A_\mu\!=\!(\pm N,0,0,0)$.\footnote
{This corresponds to normalization of the scalar potential to $1$ rather than to $0$ at spatial infinity.}
This solution is thus {\em not} vacuum, in contrast to the other three rings we include for comparison.

We will consider a special case of the Majumdar-Papapetrou (MP) configuration, consisting of a number of identical extreme black holes arranged in a circle, actually a continuous limit of this situation when one has an infinite number of ``infinitesimal'' holes distributed along a circular ring. Even without referring to the black-hole uniqueness theorems, it is to be expected that such a ring will be singular, although resembling a horizon in a sense that the lapse function will be zero on it. Namely, in making the limit, the individual horizons become infinitesimal, so the curvature on them diverge necessarily, in agreement with conclusion of \cite{HartleH-72} (section III there).
The system described has already been studied by \cite{JaramilloL-11} who solved numerically the Bishop's equations in order to localize its apparent horizon. We plan to revisit this question in a next paper (and compare the results for the different rings considered here).

Denoting the Weyl radius of the ring by $a$, the lapse is given by
\begin{align}
  \frac{1}{N}
  & =1+\frac{M}{2\pi}\int\limits_0^{2\pi}\frac{{\rm d}\phi'}{\sqrt{\rho^2+a^2-2a\rho\cos(\phi\!-\!\phi')+z^2}}=
  \nonumber \\
  & =1+\frac{2MK(k)}{\pi l_2} \;,  \label{1/N}
\end{align}
where $M$ stands for the ring's total mass,
\[l_{1,2}:=\sqrt{(\rho\mp a)^2+z^2}\]
and
\[K(k) := \int_0^{\pi/2}\frac{{\rm d}\alpha}{\sqrt{1-k^2\sin^2\alpha}}\]
is the complete elliptic integral of the first kind, with modulus and complementary modulus
\[k^2:=1-\frac{(l_1)^2}{(l_2)^2}=\frac{4a\rho}{(l_2)^2}\;, \qquad
  k'^2:=1-k^2=\frac{(l_1)^2}{(l_2)^2} \;.\]
Note that the second term of (\ref{1/N}) represents the (minus) ``Newtonian'' potential of a ring and (thus) also appears in the Bach-Weyl ring solution which will be discussed in a next section.

Especially on the axis ($\rho\!=\!0$) and in the equatorial plane ($z\!=\!0$), one has
\begin{align*}
  & \rho=0: \quad k=0, \; K=\pi/2 \quad \Rightarrow \quad
    \frac{1}{N}=1+\frac{M}{\sqrt{z^2+a^2}} \,, \\
  & z=0: \quad l_2=\rho+a \quad \Rightarrow \quad
    \frac{1}{N}=1+\frac{2MK\!\left(\frac{2\sqrt{a\rho}}{\rho+a}\right)}{\pi\,(\rho+a)}\,.
\end{align*}
At radial infinity ($\rho^2+z^2\rightarrow\infty$) the elliptic-integral term vanishes, the lapse approaches unity and the metric becomes flat. In a vicinity of the ring ($l_1\rightarrow 0^+$, $l_2\rightarrow 2a^-$, $k\rightarrow 1^-$), on the other hand, the elliptic integral is very large and the lapse goes to zero as
\[N(k\!\rightarrow\!1^-)\sim\frac{\pi l_2}{2MK(k)}\sim\frac{\pi a}{M\ln(8a/l_1)} \;.\]

In toroidal coordinates, the important functions read
\[(l_{1,2})^2=2a^2\,\frac{\cosh\zeta\mp\sinh\zeta}{\cosh\zeta-\cos\psi}
             =\frac{2a^2\,e^{\mp\zeta}}{\cosh\zeta-\cos\psi} \,,\]
\[k^2=1-e^{-2\zeta}, \quad k'^2=e^{-2\zeta}.\]

In the $a\!\rightarrow\!0^+$ limit, $k\!=\!0$, so $K(k)\!=\!\pi/2$ and the lapse function reduces to
\[\frac{1}{N}=1+\frac{M}{\sqrt{\rho^2+z^2}} \,.\]
The ellipsoidal coordinates $(R,\vartheta)$ become spherical (with respect to the Weyl axes) and the metric reduces to
\[{\rm d}s^2=-\frac{{\rm d}t^2}{\left(1\!+\!\frac{M}{R}\right)^{\!2}}\,
             +\left(\!1\!+\!\frac{M}{R}\right)^{\!\!2} \!
              \left[{\rm d}R^2\!+\!R^2({\rm d}\vartheta^2\!+\!\sin^2\vartheta\,{\rm d}\phi^2)\right]\]
which is the extreme Reissner-Nordstr\"om metric in isotropic coordinates.
Interestingly, the central location $R\!=\!0$ where the ring has shrunk then represents an extreme horizon. (This is clearly not a simple limit, but it may have been expected, as a return to the original, ``point-like'' Majumdar-Papapetrou--type source. More specifically, it well illustrates the arguments given in section III of \cite{HartleH-72}.)

\subsection{Bach-Weyl (BW) ring}

What may be considered the most ``ordinary'' ring is described by the Bach-Weyl solution \cite{Bach-Weyl-22},
\begin{align}
  \frac{1}{N} &=\exp\frac{2MK(k)}{\pi l_2} \;, \\
  \lambda     &= -\frac{M^2}{4\pi^2 a^2\rho} \times \nonumber \\
              & \quad\times
                \left[(\rho\!+\!a)(E\!-\!K)^2+\frac{(\rho-a)(E\!-\!k'^2 K)^2}{k'^2}\right],
                \label{lambda-BW}
\end{align}
where $M$ and $a$ again denote the ring's mass and Weyl radius,
\[E\equiv E(k) := \int_0^{\pi/2}\sqrt{1-k^2\sin^2\alpha}\;{\rm d}\alpha\]
is the complete elliptic integral of the second kind,
and $l_{1,2}$, $k$ and $k'$ are defined as in the MP-ring case.
Hence, the MP ring can be viewed as a certain approximation of the BW ring for $MK(k)\!\ll\!l_2$.
However, as opposed to the MP-ring solution, the Bach-Weyl solution is also described, in addition to $N$, by the ``second'' metric function $\lambda$ which has no Newtonian analogue; this vanishes on the axis, but is important close to the ring. There, the behaviour of $\lambda$ makes the space properties very different from those of the Newtonian case. In particular, the space is strongly anisotropic rather than locally cylindrical around the ring \cite{Hoenselaers-95}, more precisely, the BW ring is not ``simple'' in the sense of Israel \cite{Israel-77} (i.e., it is not a ``normal-dominated'' singularity as defined and treated by \cite{Liang-73}).

In comparison with the Majumdar-Papapetrou case, for the Bach-Weyl solution the lapse vanishes at the ring faster, like
\[N(k\!\rightarrow\!1^-)\sim\exp\!\left(-\frac{2MK(k)}{\pi l_2}\right)
                     \sim \left(\frac{l_1}{8a}\right)^{\!M/(\pi a)}.\]

Finally, in the $a\!\rightarrow\!0^+$ limit, both elliptic integrals yield $\pi/2$ and the metric functions reduce to
\[\frac{1}{N}=\exp\frac{M}{\sqrt{\rho^2+z^2}} \,, \qquad
  \lambda=-\frac{1}{2}\left(\frac{M\rho}{\rho^2+z^2}\right)^{\!2},\]
which is a Curzon solution (it appears as a directional point-like singularity in Weyl coordinates, but in fact keeps a non-trivial, ring-like structure, as unveiled in detail by \cite{ScottS-86}).

\subsection{Appell ring}

The simplest of the Appell-ring metrics is given by $N\!\equiv\!e^\nu$ and $e^\lambda$ with
\begin{align}
 \nu &= \mp\frac{M}{\sqrt{2}\,\Sigma}\,\sqrt{\Sigma+\rho^2+z^2-a^2}  \label{nu_Appell} \\
     &= \mp\frac{M}{\sqrt{2\Sigma}}\,(1+\cos\psi) \\
     &= -\frac{MR}{\Sigma} \;,  \label{nuApp} \\
 \lambda
     &= \frac{M^2}{8a^2}
        \left[1\!-\frac{\rho^2\!+\!z^2\!+\!a^2}{\Sigma}
               -\frac{2a^2\rho^2(\Sigma^2\!-\!8z^2 a^2)}{\Sigma^4}\right] \\
     &= -\frac{M^2}{16a^2}\,(2\cosh\zeta-2+\sinh^2\zeta\,\cos 2\psi) \\
     &= -\frac{M^2\sin^2\vartheta}{4\,\Sigma}\!
        \left[1\!+\frac{(R^2\!+\!a^2)(\Sigma^2\!-\!8R^2 a^2\cos^2\vartheta)}{\Sigma^3}\right]\!,
        \label{lambdaApp}
\end{align}
where $M$ and $a$ again denote the ring mass and Weyl radius, $\Sigma$ is given by (\ref{Sigma}), and the first/second/third expressions are in the Weyl/toroidal/oblate coordinates (described in section \ref{metric,coords}).

General relativistic space-times generated by Appell rings have been mainly analyzed by \cite{GleiserP-89} (this kind of sources appeared in electrostatics originally). As discussed and illustrated in \cite{SemerakZZ-99a} (Appendix A there), the spatial structure of the Appell solution is similar to that of the Kerr solution (where, however, $\rho$ and $z$ must be taken the Kerr-Schild cylindrical coordinates rather than the Weyl ones), but a horizon and rotational dragging are not present, naturally. In particular, both space-times have the disc $(z\!=\!0,\rho\!\leq\!a)$ $\Leftrightarrow$ $R\!=\!0$ at their centre, which is intrinsically flat but whose ring-like boundary $[z\!=\!0,\rho\!=\!a]$ $\Leftrightarrow$ $[R\!=\!0,\vartheta\!=\!\pi/2]$ represents a curvature singularity ($\Sigma\!=\!0$). If approaching the disc from either side (along $\vartheta\neq\pi/2$), $R$ decreases to zero, whereas its gradient does not vanish, which indicates that the manifold continues, across the disc serving as a branch cut, smoothly to the second asymptotically flat sheet characterized by $R\!<\!0$; there, (\ref{nu_Appell}) should be taken with the bottom sign.\footnote
{Keeping the right sign in formulas is at times annoying, mainly in terms containing odd powers of $z$, because both space-time sheets are reflection symmetric, while, at the same time, $z$ switches sign across both the equatorial planes as well as across the $R\!=\!0$ circle. It helps to realize that {\em physically} the two sheets differ in the sign of $M$: hence, the only safe way how to cover all the possibilities for such terms like $Mz$ is to write them as $M|z|$ and to change sign of $M$ in the second sheet. The same remark also applies to the Kerr solution below if using the Kerr-Schild coordinates. Clearly it is more comfortable to use the oblate coordinates, because $R$ (or $r$) in itself ensures a correct sign without any caution (and without adjusting the sign at $M$).}
Unless admitting the second sheet, a layer of mass would be present on the $R\!=\!0$ disc whose ``Newtonian'' surface density $w(\rho)$ can be found by using the potential (\ref{nu_Appell}) in the relation
\[\lim\limits_{z\rightarrow 0^+}\nu_{,z}=2\pi w(\rho)\]
valid in Weyl space-times:
\[w = -\frac{Ma}{2\pi\,(a^2-\rho^2)^{3/2}}
    = -\frac{M}{2\pi a^2\cos^3\vartheta} \;.\]
This is everywhere negative and even diverging to $-\infty$ toward the disc edge, while finally jumping to $+\infty$ at the very singular rim (to ensure the finite positive total mass $M$).
Irrespectively of the interpretation, in the spherical region $\rho^2+z^2<a^2$ $\Leftrightarrow$ $0\!\leq\!R\!<\!a|\cos\vartheta|$ the field is ``repulsive" in the sense that momentarily static particles are accelerated {\em away} from the central disc.

Clearly $\nu$ is everywhere negative, it only vanishes in the interior of the Appell ring (on $R\!=\!0$); it is particularly simple in the equatorial plane outside of the ring ($\vartheta\!=\!\pi/2$), $\nu\!=\!-M/R$. The second function $\lambda$ is almost everywhere negative too, but it is also positive in a certain small region, namely -- within the $(\rho,z)$ quarter-plane -- in a crescent around $R^2\!=\!a^2\cos^2\vartheta$ (which is just the boundary of the ``repulsive'' region). Actually, on this sphere one has $\Sigma\!=\!2R^2\!=\!2a^2\cos^2\vartheta$, which yields
\[\lambda=\left(\frac{M}{4a}\,\tan^2\vartheta\right)^{\!2}.\]
In the equatorial plane ($\vartheta\!=\!\pi/2$), $\lambda$ reduces to
\[\lambda=-\frac{M^2(2R^2+a^2)}{4R^4} \;,\]
while in the ring's interior ($R\!=\!0$) it reads
\[\lambda=-\left(\frac{M\sin\vartheta}{2a\cos^2\vartheta}\right)^{\!2}(1+\cos^2\vartheta) \,.\]

In the $a\!\rightarrow\!0^+$ limit, $\Sigma\!=\!\rho^2+z^2\!=\!R^2$ and the metric functions reduce to
\begin{align*}
  \nu &= \mp\frac{M}{\sqrt{\rho^2+z^2}}=-\frac{M}{R} \,, \\
  \lambda &= -\frac{1}{2}\left(\frac{M\rho}{\rho^2+z^2}\right)^{\!2}
           = -\frac{M^2\sin^2\vartheta}{2R^2} \,,
\end{align*}
so one is left, as in the BW-ring case, with the Curzon solution.

Several references to the literature should be added at this place.
Firstly, Zipoy \cite{Zipoy-66} showed that solving the static axisymmetric vacuum problem in oblate spheroidal coordinates leads to a ring-like singularity in the equatorial plane, whereas in prolate coordinates a finite line singularity typically arises along the symmetry axis.\footnote
{Later, within studies of larger classes of Weyl-type solutions, even some prolate-type metrics turned out to possess singularities which are geometrically ring-like -- see \cite{KodamaH-03}.}
In particular, his monopole oblate solution (equation (17) in \cite{Zipoy-66}) is described by the same metric as the Appell ring, only with a different potential
\[\nu=-\frac{M}{a}\,{\rm arctan}\frac{a}{R}\,.\]
In interpreting the metric, Zipoy also arrived at a double-sheeted topology, connected through the $R\!=\!0$ circle.
An alternative interpretation -- the one involving just one sheet and generated instead by a surface layer of mass present on the circle $z\!=\!0$, $\rho\!\leq\!a$ -- was then suggested by \cite{BonnorS-68}; the Newtonian density necessary to produce the above potential is positive,
\[w = \frac{\nu_{,z}(z\!\rightarrow\!0^+)}{2\pi}
    = \frac{M}{2\pi a^2\cos\vartheta} \,,\]
hence less unphysical than that corresponding to the Appell-ring solution.

Note also that the rings through which double-sheeted solutions are connected in fact serve as wormholes, and they are indeed being analyzed in this context -- see \cite{GibbonsV-16}, for example.

\subsection{Kerr ring}

The Kerr metric is mostly being given in the Boyer-Lindquist coordinates $(t,r,\theta,\phi)$,
\begin{equation}
  {\rm d}s^2
    =-N^2\,{\rm d}t^2
     +g_{\phi\phi}\,({\rm d}\phi\!-\!\omega\,{\rm d}t)^2
     +\frac{\Sigma}{\Delta}\,{\rm d}r^2
     +\Sigma\,{\rm d}\theta^2,
  \label{Kerr-BL}
\end{equation}
where
\begin{align*}
  & N^2 = \frac{\Sigma\Delta}{{\cal A}}=1-\frac{2Mr(r^2+a^2)}{\cal A} \;, \\
  & g_{\phi\phi} = \frac{{\cal A}}{\Sigma}\sin^2\theta \;, \qquad 
    \omega := \frac{-g_{t\phi}}{g_{\phi\phi}}=\frac{2Mar}{{\cal A}} \;, \\
  & \Sigma := r^2+a^2\cos^2\theta, \qquad
    \Delta := r^2-2Mr+a^2, \\
  &{\cal A} := (r^2+a^2)^2-\Delta a^2\sin^2\theta \\  &\quad
             =\Sigma(r^2+a^2)+2Mra^2\sin^2\theta  \\  &\quad
             =\Sigma\Delta+2Mr(r^2+a^2) \,.
\end{align*}

The basic properties of the Kerr solution need not be reminded, just that $\Sigma\!=\!0$ gives the singularity, while elsewhere $\Sigma$ is positive; $\Delta\!=\!0$ gives the horizons, being negative between them and positive everywhere else; like for the Appell ring, $N\!=\!1$ on the central circle given by $r\!=\!0$; between the horizons, $N^2\!<\!0$, while elsewhere it is positive; and $\omega$ is positive/negative on the $r\!>\!0$/$r\!<\!0$ sheet, with smooth zero on $r\!=\!0$. However, there is one exception to the last two properties: $N^2$ is {\em negative} and $\omega$ is {\em positive} where ${\cal A}\!<\!0$; the latter holds in a toroidal region spanned by the singularity and lying entirely in the $r\!<\!0$ space (closed time-like loops exist in this peculiar region, because $g_{\phi\phi}\!<\!0$ there).

$M$ denotes mass and $a$ is the radius of the singular ring in the cylindrical Kerr-Schild coordinates $(T,\rho,z,\psi)$ which are related to the Boyer-Lindquist ones by
\begin{align*}
  &{\rm d}T={\rm d}t-\frac{2Mr}{\Delta}\,{\rm d}r \,, \qquad
   {\rm d}\psi={\rm d}\phi-\frac{2Mar}{(r^2+a^2)\Delta}\,{\rm d}r \,, \\
  &\quad \rho=\sqrt{r^2+a^2}\,\sin\theta, \qquad z=r\cos\theta
\end{align*}
and in which the metric assumes the form
\begin{align}
  {\rm d}s^2
  = & -{\rm d}T^2+{\rm d}\rho^2+\rho^2{\rm d}\psi^2+{\rm d}z^2+ {} \nonumber \\
    & +\frac{2Mr^3}{r^4+a^2 z^2}
       \left({\rm d}T+\frac{r\rho\,{\rm d}\rho-a\rho^2{\rm d}\psi}{r^2+a^2}+\frac{z{\rm d}z}{r}\right)^2 \!\!.
  \label{Kerr-Schild}
\end{align}
The oblate radius $r$ satisfies the equation
\[r^4-(\rho^2-a^2+z^2)\,r^2-a^2 z^2=0 \,.\]
Clearly the meridional-plane transformation $(r,\theta)\!\leftrightarrow\!(\rho,z)$ is the same as the one between the Weyl and the oblate spheroidal coordinates (\ref{oblate}) which we employed for the Appell ring. And, actually, the ``monopole'' part of the Kerr field represented in the cylindrical Kerr-Schild coordinates is very similar to the Appell field represented in the Weyl coordinates \cite{SemerakZZ-99a} and we thus use the same notation $(\rho,z)$ in this paper.\footnote
{The Weyl coordinates, on the other hand, are not useful for the innermost parts of the Kerr space-time, since in them the outer horizon is mapped onto a finite segment of the axis ($\rho\!=\!0$, $|z|\leq\sqrt{M^2\!-\!a^2}$) and its interior is not covered at all.}
(However, the Boyer-Lindquist coordinates are denoted by usual $r$, $\theta$, i.e. differently from the oblate ones $R$, $\vartheta$ used in static-ring fields, as already stressed in section \ref{metric,coords}.)

Although we include the Kerr ring into this comparison, it should be stressed again that it is a source of a more general, stationary but non-static space-time, namely it bears a non-zero angular momentum (given by $Ma$) and thus drags the space around in differential co-rotation. This implies, among others, that it is not so clear what to take as relevant ``meridional plane''. The latter is not important for {\em plotting} of the quantities, because of the axial symmetry, but it will be important when some quantity shall really describe the chosen two-dimensional ``meridional'' surface (such as the Gauss curvature, for example).
Another difference is that for $a\!\leq\!M$ the Kerr ring is contained within a stationary black hole, i.e. it is surrounded by a marginally trapped surface whose history represents a Killing horizon.

\subsection{On illustrations and the double-sheeted topology of the Appell-ring and Kerr space-times}

Below, we check some basic properties of the above ring fields. In illustrating them, we plot the behaviour within the meridional plane $\{t\!=\!{\rm const},\phi\!=\!{\rm const}\}$ which is privileged by the Killing symmetries and usually provides the best insight. The plots are given in natural coordinates $(\rho,z)$ which for the static (MP, BW and Appell) rings are the Weyl coordinates, while for the Kerr case they represent Kerr-Schild coordinates. As already reminded above, the Appell-ring and the Kerr space-times are double-sheeted, the two sheets being effectively distinguished by a sign of the mass $M$ and connected smoothly on the circle spanned by the ring. We include both sheets in one plots, specifically by showing the $M\!>\!0$/$M\!<\!0$ regions above/below their $z\!=\!0$ planes, so the plot is smooth across this central circle, whereas ``outside of the rings'' (at $\rho\!>\!a$) the equatorial planes do not match. (However, if some quantity behaves in the same manner in both sheets, the equatorial planes {\em visually} do match in its plot.)

Note also that most of the simple-property figures show rings with $M\!=\!a$; for the Kerr solution, this means the extreme case with double degenerate horizon (with vanishing surface gravity) found on $r\!=\!M$.

\begin{figure*}
\centering
\includegraphics[width=0.9\textwidth]{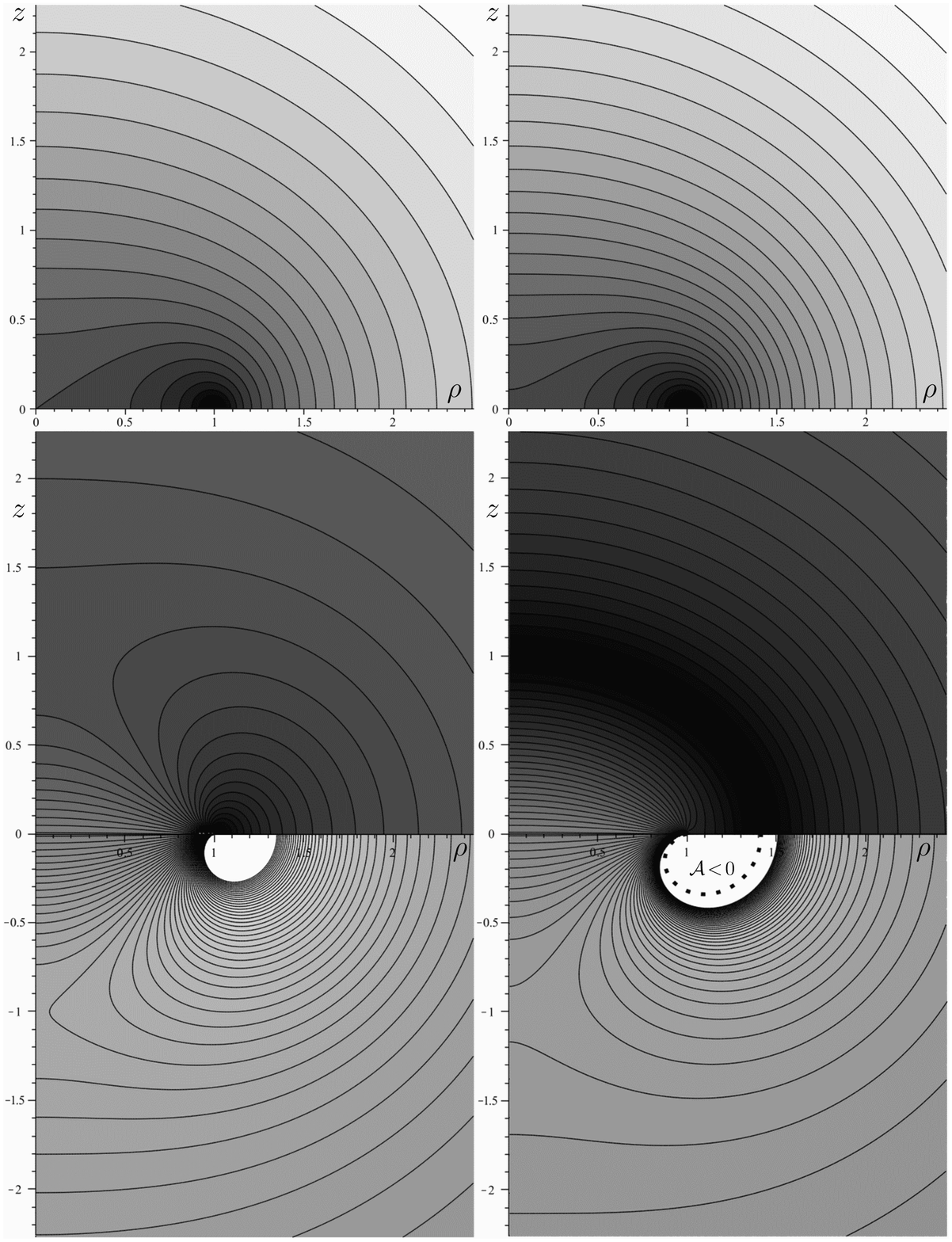}
\caption
{Contours of the lapse function $N$, plotted within the meridional plane for the MP ring (top left), the BW ring (top right), the Appell ring (bottom left) and the Kerr ring (bottom right). All the rings have $M\!=\!a$. The $(\rho,z)$ axes (scaled by $a$) represent Weyl coordinates for the static rings, while Kerr-Schild coordinates for the Kerr ring. The lapse scalar is everywhere positive, only vanishing at the very rings and on the Kerr horizon; light/dark shading indicates larger/smaller values. The contour values range from $0.4$ to $0.76$ for the MP ring, from $0.23$ to $0.73$ for the BW ring, and from $0.03$ to $3.01$ for the Appell and Kerr rings (with $N\!=\!1$ on the $z\!=\!0$ disc encircled by the ring).}
\label{lapse}
\end{figure*}

Figure \ref{lapse} shows contours of the lapse invariant $N$ which, for example, determines redshift between local static observers and asymptotic inertial ones. As expected, this ``Newtonian'' part of the field behaves quite intuitively around all the four rings. However, already at this level some peculiar features of the double-sheeted solutions appear, mainly the very high values in the negative-sheet vicinity of these two rings (white areas); in case of the Kerr ring, $N$ even diverges there on a certain toroidal surface (dotted line in the plot) inside which $N^2$ turns negative due to ${\cal A}\!<\!0$. At the positive-$r$ sheet of the Kerr field, one notices the horizon given by $N\!=\!0$.

\section{Simple properties of the ring fields}
\label{simple-properties}

\subsection{Small circumference}

We will calculate several simple sizes of the ring sources.
First, their ``small'' circumference is best calculated in toroidal coordinates $(\zeta,\psi)$, as a limit ($\zeta\!\rightarrow\!\infty$: the ring) of the integral
\begin{equation}  \label{cross-circumference}
  2\int\limits_0^\pi \sqrt{g_{\psi\psi}}\,{\rm d}\psi
  = 2a\int\limits_0^\pi \frac{e^\lambda\;{\rm d}\psi}{N\,(\cosh\zeta-\cos\psi)} \;.
\end{equation}
Note that the integrand is always positive in that limit (provided that $N\!\equiv\!e^\nu$ is such), namely behaving like $2e^{\lambda-\nu-\zeta}$.

For the MP ring ($\lambda\!=\!0$), the cross circumference of a generic $\zeta\!=\!{\rm const}$ tube amounts to
\begin{align*}
  &\int\limits_0^\pi
     \left[\frac{2a}{\cosh\zeta-\cos\psi}+
           \frac{2\sqrt{2}\,MK(\sqrt{1-e^{-2\zeta}})}{\pi\,e^{\zeta/2}\,\sqrt{\cosh\zeta-\cos\psi}}
           \right]
     {\rm d}\psi= \\
  & =\frac{2\pi a}{\sinh\zeta}+
     \frac{4\sqrt{2}\,MK(\sqrt{1-e^{-2\zeta}})}{\pi\,e^{\zeta/2}\,\sqrt{\cosh\zeta+1}}\,
     K\!\left(\!\frac{\sqrt{2}}{\sqrt{\cosh\zeta+1}}\!\right).
\end{align*}
In the limit $\zeta\!\rightarrow\!\infty$, both terms vanish.
Hence, although the ring is the place where the lapse $N$ vanishes, so it corresponds to a horizon of the usual Weyl solutions, it has zero cross circumference. (Only in the $a\!\rightarrow\!0^+$ limit we saw that it really turns into an extreme horizon.)

For the BW ring, the integrand of (\ref{cross-circumference}) behaves, in the $\zeta\!\rightarrow\!\infty$ limit, as
\begin{equation}  \label{g_psipsi,BW}
  2\,e^{\lambda-\nu-\zeta}
  \sim 2\exp\!\left(-\frac{2M^2\cos\psi}{\pi^2 a^2}\,\zeta^2 e^\zeta\right),
\end{equation}
which means that it extremely strongly vanishes on the $\cos\psi\!>\!0$ side (when the ring is approached from the ``outer'' half-space, i.e. from $\rho\!>\!a$), whereas it extremely strongly diverges on the $\cos\psi\!<\!0$ side (when the ring is approached from the ``inner'' half-space, i.e. from $\rho\!<\!a$). For $\cos\psi\!=\!0$ (i.e. from directions perpendicular to the ring's plane), the limit also vanishes,
\begin{equation}  \label{g_psipsi,BW,Pi/2}
  2\,e^{\lambda-\nu-\zeta}(\psi\!=\!\pm\pi/2)
  \sim 2\exp\!\left(-\frac{5M^2\zeta^2}{\pi^2 a^2}\right).
\end{equation}
Hence, the outer half-circumference vanishes, while the inner half-circumference (and thus also the sum of both) {\em diverges}.

For the Appell ring, the integrand of (\ref{cross-circumference}) goes like
\begin{equation}  \label{g_psipsi,Appell}
  2\,e^{\lambda-\nu-\zeta}
  \sim 2\exp\!\left(-\frac{M^2\cos 2\psi}{64a^2}\,e^{2\zeta}\right),
\end{equation}
so it vanishes/diverges at the ring -- even stronger than for the BW ring -- from the directions $\cos 2\psi\!>\!0\,$/$\,\cos 2\psi\!<\!0$. Especially for $\cos 2\psi\!=\!0$ (i.e. $\psi\!=\!n\cdot\pi/4$) it vanishes as well, according to
\begin{equation}  \label{g_psipsi,Appell,Pi/2}
  2\,e^{\lambda-\nu-\zeta}
  \sim 2\exp\!\left(-\frac{M^2}{16a^2}\,e^{\zeta}\right).
\end{equation}
The total circumference includes integrals of strongly infinite quantity over finite intervals, so it is clearly {\em infinite} like for the BW ring. Notice also that the asymptotic behaviour involves $M^2$, so it is the same for both the $R\!>\!0$ and $R\!<\!0$ (which effectively means $M\!<\!0$) sheets.

For the Kerr ring, we will stay in Boyer-Lindquist coordinates in which the $\{t\!=\!{\rm const},\phi\!=\!{\rm const}\}$ plane is described by
\[{\rm d}s^2=\frac{\Sigma}{\Delta}\,{\rm d}r^2+\Sigma\,{\rm d}\theta^2.\]
Since $\Sigma$ vanishes at the ring, its small circumference (calculated within this plane) surely does the same.

Hence, the MP and the Kerr rings give expectable (zero) results, whereas the BW and the Appell rings behave more strangely (and very differently from different directions). Let us refer to figure \ref{Kretschmann-detail} in advance, showing contours of the Kretschmann scalar in the rings' vicinity, because it nicely illustrates what have been found, including the important local angles $\psi\!=\!n\cdot\pi/2$ (for the BW ring) and $\psi\!=\!n\cdot\pi/4$ (for the Appell ring).

\subsection{Large circumference}

For the ``large'' circumference, the important quantity is the proper length of the $\{t\!=\!{\rm const},z\!=\!{\rm const},\rho\!=\!{\rm const}\}$ circles,
\begin{equation}
  2\pi\sqrt{g_{\phi\phi}}
  =\frac{2\pi\rho}{N}=2\pi\rho\,e^{-\nu}
  =2\pi\,\frac{a\,e^{-\nu}\sinh\zeta}{\cosh\zeta-\cos\psi} \;,
\end{equation}
which is then limited to the ring (the last, toroidal expression enables one to make this limit from any direction properly).

The circumferential radius $\sqrt{g_{\phi\phi}}$ of the MP ring is infinite,
\begin{align*}
  & \lim\limits_{\zeta\rightarrow\infty}\sqrt{g_{\phi\phi}}
    =\lim\limits_{\zeta\rightarrow\infty}\frac{a\,\sinh\zeta}{N(\cosh\zeta-\cos\psi)}= \\
  & =\lim\limits_{\zeta\rightarrow\infty}
     \left[\frac{a\,\sinh\zeta}{\cosh\zeta\!-\!\cos\psi}+
           \frac{\sqrt{2}\,MK(\sqrt{1\!-\!e^{-2\zeta}})\,\sinh\zeta}
                {\pi\,e^{\zeta/2}\,\sqrt{\cosh\zeta\!-\!\cos\psi}}\right]= \\
  & =a+\frac{M}{\pi}\,\lim\limits_{\zeta\rightarrow\infty}K(\sqrt{1-e^{-2\zeta}})
    =\infty \,.
\end{align*}
In the $a\!\rightarrow\!0^+$ limit, the transformation to toroidal coordinates (\ref{toroidal-coords}) does not have sense and neither the above calculation; since the ring becomes an extreme Reissner-Nordstr\"om horizon in that limit, we know the circumference is $2\pi M$ then.

For the BW ring, the circumferential radius reads
\begin{align*}
  &\lim\limits_{\zeta\rightarrow\infty}\rho\,\exp\frac{2MK(k)}{\pi l_2}=  \\
  &=\lim\limits_{\zeta\rightarrow\infty}\,
    \frac{a\,\sinh\zeta}{\cosh\zeta\!-\!\cos\psi}\;
    \exp\frac{\sqrt{2}MK(\sqrt{1\!-\!e^{-2\zeta}})}
             {\pi a\,\frac{\exp(\zeta/2)}{\sqrt{\cosh\zeta-\cos\psi}}}=  \\
  &=a\lim\limits_{\zeta\rightarrow\infty}
     \exp\frac{MK(\sqrt{1\!-\!e^{-2\zeta}})}{\pi a}
   =\infty \,.
\end{align*}
Clearly the divergence is exponentially stronger than in the above MP case.
In the $a\!\rightarrow\!0^+$ limit, one can use the spheroidal (then in fact spherical) coordinates $(R,\vartheta)$ in which
\[\sqrt{g_{\phi\phi}}=R\sin\vartheta\,\exp\frac{M}{R} \,,\]
so the circumference ($R\!\rightarrow\!0^+$) even then diverges.

The Appell-ring circumferential radius, taken from the $R\!>\!0$ sheet, is given by
\begin{align*}
  &\lim\limits_{\zeta\rightarrow\infty}\rho\,\exp\frac{M(1+\cos\psi)}{\sqrt{2\Sigma}}=  \\
  &=\lim\limits_{\zeta\rightarrow\infty}
    \frac{a\,\sinh\zeta\,
          \exp\!\left[\frac{M}{2a}(1\!+\!\cos\psi)\sqrt{\cosh\zeta\!-\!\cos\psi}\right]}
         {\cosh\zeta\!-\!\cos\psi}=  \\
  &=a\lim\limits_{\zeta\rightarrow\infty}
     \exp\!\left[\frac{Me^{\zeta/2}}{2\sqrt{2}\,a}(1\!+\!\cos\psi)\right]
   =\infty \,,
\end{align*}
only for $\psi\!=\!\pi$ (i.e. if the circumference is measured {\em exactly} from inside of the ring, in the ring's plane) the circumferential radius is just $a$. In the $a\!\rightarrow\!0^+$ limit, one gets the Curzon ``particle'' like with the BW ring, so the circumference diverges.

For the Kerr ring, we will again stay in Boyer-Lindquist coordinates, because there $\sqrt{\Sigma}$ represents ``distance'' from the ring that may be taken along any direction. We have
\[\lim\limits_{\Sigma\rightarrow 0}\sqrt{g_{\phi\phi}}
   =\lim\limits_{\Sigma\rightarrow 0}\sqrt{\frac{\cal A}{\Sigma}}\,\sin\theta
   =a\lim\limits_{\Sigma\rightarrow 0}\sqrt{1+\frac{2Mr}{\Sigma}} \;,\]
which is only finite -- equal to $a$ -- if taken exactly from the $r\!=\!0$ direction (thus from inside of the ring), whereas from all other local latitudinal directions it diverges.
In the $a\!\rightarrow\!0^+$ (Schwarzschild) limit, the circumference vanishes, independently of direction (of course).

\subsection{Proper radius, proper distances}

Proper radius of the rings is calculated by
\begin{equation}
  \int\limits_0^a \sqrt{g_{\rho\rho}(z\!=\!0)}\;{\rm d}\rho
  =\int\limits_0^a \left(\frac{e^\lambda}{N}\right)_{\!z=0}\,{\rm d}\rho \;.
\end{equation}
If wanting to find a proper distance to the ring from a {\em generic} direction, it is again suitable to perform it in toroidal coordinates,
\begin{equation}  \label{distance-to-ring}
  \int\limits_\zeta^\infty \sqrt{g_{\zeta\zeta}(\psi)}\;{\rm d}\zeta
  =a\int\limits_\zeta^\infty \frac{e^\lambda\,{\rm d}\zeta}{N(\cosh\zeta-\cos\psi)} \;.
\end{equation}
Note that $g_{\zeta\zeta}\!=\!g_{\psi\psi}$, so the integrand is the same as that for the cross circumference, expression (\ref{cross-circumference}).

For the MP ring this distance is finite from all directions ($\psi$), since
\[\int\limits_\zeta^\infty
     \left[\frac{a}{\cosh\zeta-\cos\psi}+
           \frac{\sqrt{2}\,MK(\sqrt{1-e^{-2\zeta}})}{\pi\,e^{\zeta/2}\,\sqrt{\cosh\zeta-\cos\psi}}
           \right]
     {\rm d}\zeta\]
has both integrands vanishing quite fast at $\zeta\!\rightarrow\!\infty$ -- the first one as $2ae^{-\zeta}$ and the second one as $(2M/\pi)\zeta e^{-\zeta}$.
In particular, the MP-ring proper radius comes out, as obtained in the Weyl coordinates,
\begin{align}
  \int\limits_0^a
    \left[1+\frac{2MK\!\left(\frac{2\sqrt{a\rho}}{\rho+a}\right)}{\pi\,(\rho+a)}\right]
    {\rm d}\rho
  & =a+\frac{2M}{\pi}\int_0^1 K(v)\,{\rm d}v=  \nonumber \\
  & =a+\frac{4{\cal G}}{\pi}M \,,  \label{MP-radius}
\end{align}
where
\[{\cal G}=\frac{1}{2}\int_0^1 K(v)\,{\rm d}v
   =\sum\limits_{l=0}^\infty\frac{(-1)^l}{(2l+1)^2}
   \doteq 0.9159656\]
is the Catalan's constant.
In the $a\!\rightarrow\!0^+$ limit, the result apparently does not work well, namely it remains finite, but we know the MP ring turns into an extremally charged spherical horizon in that limit, and such a horizon is at {\em infinite} proper radial distance from all directions.

For the BW ring the integrand of (\ref{distance-to-ring}) behaves, for $\zeta\!\rightarrow\!\infty$, as
$2\,e^{\lambda-\nu-\zeta}$. Regarding the limit behaviour of this term given by (\ref{g_psipsi,BW}) and (\ref{g_psipsi,BW,Pi/2}), we see that the distance measured from any direction on the $\cos\psi\geq 0$ side (from the outer half-plane, $\rho\geq a$) is finite, whereas the distance from the $\cos\psi\!<\!0$ side (from the inner half-plane, $\rho\!<\!a$) strongly diverges. This means, in particular, that the proper radius of the BW ring is infinite.
In the $a\!\rightarrow\!0^+$ limit, when the BW ring shrinks to a Curzon ``particle'', toroidal coordinates are not meaningful and one rather uses the spheroidal/spherical ones. In these,
\[\sqrt{g_{RR}}=\frac{e^\lambda}{N}
               =\frac{\exp\frac{M}{R}}{\exp\frac{M^2\sin^2\vartheta}{2R^2}}\]
in that limit, which yields finite proper distance from any direction except along the axis ($\sin\vartheta\!=\!0$).

For the Appell solution, the $2e^{\lambda-\nu-\zeta}$ term behaves like (\ref{g_psipsi,Appell}) and (\ref{g_psipsi,Appell,Pi/2}), so the distance to the ring is finite from the directions $\cos 2\psi\!\geq\!0$ whereas infinite from $\cos 2\psi\!<\!0$. This remains true if the ring is shrunken to $a\!=\!0$. It means, in particular, that the ring is at finite distance when approached along the equatorial plane ($\psi\!=\!0,\pi$), whereas it is infinitely remote from perpendicular directions ($\psi\!=\!\pm\pi/2$). Since for $z\!=\!0$, $\rho\!<\!0$ (i.e. $R\!=\!0$ in short) we have
\[\nu=0 \qquad {\rm and} \qquad
  \lambda=-\frac{M^2\rho^2(2a^2-\rho^2)}{4a^2(a^2-\rho^2)^2} \;,\]
the proper radius of the ring is, as expressed in the Weyl coordinates,
\begin{align}
  & \exp\frac{M^2}{4a^2}
    \int_0^a\exp\!\left[-\frac{M^2 a^2}{4(a^2-\rho^2)^2}\right]{\rm d}\rho=  \nonumber \\
  &=\sqrt{\frac{\pi}{8}}\;a\;\exp\frac{M^2}{4a^2}\;\,
    G^{3,0}_{2,3}
    {\small
    \left(\!\!\left.\begin{array}{c}3/4,\;5/4\\ 0,\;1/2,\;1\end{array}\right|\frac{M^2}{4a^2}\right)},
    \label{area-Appell}
\end{align}
where the second expression uses the Meijer $G$-function.
For any given $a$, this decreases monotonically from $a$ to zero with $M$ increased from zero to infinity. The $a\!\rightarrow\!0^+$, Curzon-solution limit is the same as for the BW ring.

The Kerr ring is clearly at finite distance from any direction due to the factor $\Sigma$ (vanishing at the ring) standing in front of ${\rm d}r^2$ as well as ${\rm d}\theta^2$ in the Kerr meridional-plane metric. In particular, the ring's proper radius is
\[\int\limits_0^{\pi/2}\sqrt{\Sigma(r\!=\!0)}\;{\rm d}\theta
  =a\int\limits_0^{\pi/2}\cos\theta\,{\rm d}\theta
  =a \,.\]

\subsection{Proper area of the enclosed circle}

The proper area of the circle enclosed by any of the rings is 
\begin{align*}
  & 2\pi\int\limits_0^a \sqrt{(g_{\rho\rho}g_{\phi\phi})_{z=0}}\;{\rm d}\rho
  = 2\pi\int\limits_0^a \rho\left(\frac{e^\lambda}{N^2}\right)_{\!z=0}\,{\rm d}\rho
  \,.
\end{align*}

For the MP ring, $\lambda\!=\!0$ and the area is finite,
\begin{align}
  & 2\pi\int\limits_0^a
     \rho\left[1+\frac{2MK\!\left(\frac{2\sqrt{a\rho}}{\rho+a}\right)}{\pi\,(\rho+a)}\right]^{\!2}
     {\rm d}\rho =  \nonumber \\
  & =\pi a^2+8Ma+\frac{14\,\zeta(3)}{\pi}\,M^2 \,,  \label{MP-area}
\end{align}
where the Riemann zeta function at $s\!=\!3$
\begin{align*}
  \zeta(3)&=\frac{4}{7}\int_0^1 v\,K^2(v)\,{\rm d}v
           =\sum\limits_{n=1}^\infty\frac{1}{n^3}
           =\frac{8}{7}\sum\limits_{l=0}^\infty\frac{1}{(2l+1)^3} \doteq \\
          &\doteq 1.2020569
\end{align*}
is known as Ap\'ery's constant.
Note that if the $a\!\rightarrow\!0^+$ limit was made naively, the area would remain non-zero and finite, namely given just by the last term $14\zeta(3)M^2/\pi\doteq 5.357M^2$, while the ring's proper radius (previous subsection) would come out $4{\cal G}M/\pi\doteq 1.166M$. Relation of these two values is not so far from Euclidean: $\pi$ times the radius squared amounts to $4.273M^2$. (However, the MP ring in fact ``shrinks'' to an extreme Reissner-Nordstr\"om horizon and it is not clear how to understand the above two quantities then.)

For the BW ring, the area is $2\pi\int_0^a \rho\,(e^{\lambda-2\nu})_{z=0}\,{\rm d}\rho$. As expected (proper radius as well as proper circumference of the BW ring diverge), it is infinite.

The area encircled by the Appell ring is finite on the contrary,
\begin{align*}
  & 2\pi\,\exp\frac{M^2}{4a^2}
    \int_0^a \rho\,\exp\!\left[-\frac{M^2 a^2}{4(a^2-\rho^2)^2}\right]{\rm d}\rho = \\
  & = \pi a^2+
      \frac{\pi^{3/2}}{2}\,Ma\left[{\rm erf}\!\left(\frac{M}{2a}\right)-1\right]\exp\frac{M^2}{4a^2} \,,
\end{align*}
namely decreasing from $\pi a^2$ to zero monotonically with $M$ increasing from zero to infinity. It also vanishes in the $a\!\rightarrow\!0^+$ limit. Comparing the area with the proper radius (\ref{area-Appell}), one finds that with $M$ going from zero to infinity, the area/radius$^2$ ratio only slightly deviates from the Euclidean value $\pi$ -- namely it increases towards $4$, as also confirmed by the $M\!\rightarrow\!\infty$ asymptotics
\[{\rm area} \sim \frac{2\pi a^4}{M^2} \;, \qquad
  ({\rm proper~radius})^2 \sim \frac{\pi a^4}{2M^2} \;.\]

Finally, the area encircled by the Kerr ring (i.e. the area of the surface given by $r\!=\!0$) is very simple,
\[2\pi\!\int\limits_0^{\pi/2}\!\!\sqrt{{\cal A}(r\!=\!0)}\,\sin\theta\,{\rm d}\theta
  =2\pi a^2\!\!\int\limits_0^{\pi/2}\!\cos\theta\sin\theta\,{\rm d}\theta
  =\pi a^2 \,,\]
so it relates to the ring's proper radius $a$ in a Euclidean way.

\subsection{Radius and area for the Appell and Kerr rings}
\label{Appell-Kerr-circle}

An important note must be added here, concerning the Appell and the Kerr solution:
we have called ``proper radius'' the integral of $\sqrt{g_{\rho\rho}}$ calculated, in the $\rho$ direction from $0$ to $a$, along the central circle lying at $z\!=\!0$, and, similarly, we have called ``proper area'' the double integral of $\sqrt{g_{\rho\rho}g_{\phi\phi}}$ over that circle. But this is a clear choice only if one adopts the single-sheet interpretation involving the layer of mass spread over that circle. Actually, in the case of the MP and BW rings, the field falls, in the radial direction, smoothly to zero when approaching the circle enclosed by the ring, so this circle is a stable equilibrium in a vertical sense and, therefore, is a geodesic surface (it is spanned by geodesics launched tangentially to it). In the Appell and Kerr cases, on the contrary, the field is non-zero in the ring's interior, pulling the test particles from the $R\!<\!0$ to the $R\!>\!0$ side, so the central circle is {\em not} a geodesic surface (geodesics starting tangentially to this surface are deflected into the $R\!>\!0$ region).

Hence, it is rather ambiguous what to call ``proper radius'' and ``proper area'' of the Appell and Kerr rings. (This issue was already pointed out by \cite{Zipoy-66} in his study of the monopole ring solution.) However, the above ``naive'' definition has lead to plausible results for both the double-sheeted rings (for Kerr it even yields $a$ for the radius and $\pi a^2$ for the area), which may serve as a certain justification.

\begin{table*}
\begin{center}
\begin{tabular}{|l|l|l|l|l|l|}
  \hline
  RING
  & small circumference & large circumference & proper distance to; proper radius & encircled area & $a\!=\!0$ limit \\
       \hline
  Majumdar-P. 
       & zero & infinite & finite from all directions & finite & extreme RN \\
       ~ & ~ & ~ & ~ & ~ & horizon \\
       \hline
  Bach-Weyl 
       & infinite & infinite & finite/infinite from $\rho\!\geq\!a\,$/$\,\rho\!<\!a$ & infinite & Curzon \\
       ~ & (on $\rho\!<\!a$ side) & ~ & ($\Rightarrow$ proper radius infinite) & ~ & singularity \\
       \hline
  Appell & infinite & finite from $\psi\!=\!\pi$ side, &
           finite/infinite from $\cos 2\psi\geq\!0\,$/$\,<\!0$ & finite & Curzon \\
       (2-sheeted) & (on $\cos 2\psi\!<\!0$ side) & infinite from elsewhere & ($\Rightarrow$ proper radius finite)
         & ~ & singularity \\
       \hline
  Kerr & zero & finite from $r\!=\!0$ side, & finite from all directions & finite & Schwarzschild \\
       (2-sheeted) & ~ & infinite from elsewhere & ~ & ~ & singularity \\
  \hline
\end{tabular}
\caption{Summary of basic measures of the four rings. Leaving aside the stationary Kerr case, the most natural (and not directional) are the properties of the Majumdar-Papapetrou ring. On the other hand, the Bach-Weyl ring has turned out weird in all the above respects (which may not have been expected, because it is the most direct counterpart of the Newtonian circular ring).}
\end{center}
\end{table*}

\subsection{Axis of symmetry}

The basic requirement for the axis is a local flatness of the horizontal planes $z\!=\!{\rm const}$ which is known to be the case if $\lambda\!=\!0$ there. Actually, the circumferential radius about the axis $\sqrt{g_{\phi\phi}}=\rho/N$ (which determines the latter's circumference in the $\phi$-direction) then exactly coincides with proper distance from the axis $\sqrt{g_{\rho\rho}}\,\rho$, which is just the Euclidean picture.

The proper distance computed along the axis is everywhere finite, being given by
$\sqrt{g_{zz}(\rho\!=\!0)}=N^{-1}(\rho\!=\!0)$
which is regular for all the rings considered here.
Specifically, we have
\begin{align*}
  \sqrt{g_{zz}(\rho\!=\!0)}
    &= 1+\frac{M}{\sqrt{z^2+a^2}}    &\; & {\rm for~MP~ring} \\ 
    &= \exp\frac{M}{\sqrt{z^2+a^2}}  &\; & {\rm for~BW~ring} \\
    &= \exp\frac{MR}{R^2+a^2}      &\; & {\rm for~Appell~ring} \\
    &= \sqrt{1+\frac{2Mr}{r^2+a^2}}  &\; & {\rm for~Kerr~ring} \,,
\end{align*}
where in the Kerr case $g_{zz}$ is understood to correspond to the Kerr-Schild coordinates.

\subsection{Equatorial plane}

For the MP and BW rings, the most important quantity is the elliptic integral $K(k)$, having a divergence at $k\!=\!1$. In the equatorial plane the modulus reads $k=2\sqrt{a\rho}/(\rho+a)$ which only reaches 1 at the very ring ($\rho\!=\!a$), while elsewhere it is smaller. For the Appell and Kerr solutions, the metric functions only diverge at the rings as well.

One can also verify that there are no mass layers distributed in the equatorial planes, neither inside nor outside the rings, by making sure that all the metric components cross smoothly these planes ($z\!=\!0$, or, $\psi\!=\!n\pi$). A simple check follows by noticing that the metric functions only involve $z$ (or $\cos\theta$ and $\sin\theta$ in the Kerr case) in even powers.

\subsection{First curvature of the ring lines}

We saw that the proper radius of the rings tends to be finite, whereas their (large) circumference rather tends to be infinite. This contrast (specifically applying to the MP ring) suggests to compute the first curvature of the rings as curves. Let us do it in the Weyl coordinates. Any $\{t\!=\!{\rm const},\rho\!=\!{\rm const},z\!=\!{\rm const}\}$ ring has the purely azimuthal unit tangent vector
\[w^\mu=(0,0,1/\sqrt{g_{\phi\phi}},0)=(0,0,N/\rho,0),\]
so the corresponding curvature (``acceleration'') square amounts to
\begin{align*}
  & g_{\mu\nu}({w^\mu}_{;\alpha}w^\alpha)({w^\nu}_{;\beta}w^\beta)
    =\frac{1}{(g_{\phi\phi})^2}\,g_{ij}{\Gamma^i}_{\phi\phi}{\Gamma^j}_{\phi\phi}= \\
  & =\frac{g^{kl}g_{\phi\phi,k}g_{\phi\phi,l}}{4\,(g_{\phi\phi})^2}
    =\frac{N^6}{4\rho^4}\left[(g_{\phi\phi,\rho})^2+(g_{\phi\phi,z})^2\right]= \\
  & =\left(N_{,\rho}-\frac{N}{\rho}\right)^{\!2}+\left(N_{,z}\right)^2.
\end{align*}
At the ring $N\!=\!0$, so one is left with just $(N_{,\rho})^2+(N_{,z})^2$.

For both the MP and BW rings the above line curvature squared has a sharp infinity at the ring, independently of direction in which the ring is approached within the meridional plane. For the Appell ring, on the contrary, it is zero from all directions but from ``inside'' ($z\!=\!0$, $\rho\!<\!a$) from where it diverges. For the Kerr metric, the same quantity, also calculable from $(1/4)(g^{kl}g_{\phi\phi,k}g_{\phi\phi,l})/(g_{\phi\phi})^2$, diverges at the ring from all directions.

\section{Geodesics and acceleration}
\label{geod-acceleration}

\subsection{Geodesic time of flight}

Besides asymptotic regions, there are only two ``absolute'' locations between which the time of flight can be compared meaningfully -- the central point of the ring (lying on the symmetry axis) and the ring itself. More precisely, this is only meaningful for the MP and BW rings, because for the other two the central circle is {\em not} a geodesic surface, so the particles would not at all follow it towards the ring (see section \ref{Appell-Kerr-circle}). For the MP and BW rings, on the other hand, the point $(\rho\!=\!0,z\!=\!0)$ is an equilibrium (albeit unstable in the $\rho$ direction), so the above time of flight is actually infinite. However, when comparing the MP and BW rings, one expects the behaviour of geodesics near the rings to be important and not the behaviour near the central point. Actually, we have observed a profound difference between the MP and BW rings just there at the ring's inside: the MP ring is at finite proper distance from there, whereas the BW ring is infinitely remote (see table I).

Consider a time-like motion first. For a four-velocity $u^\mu:=\frac{{\rm d}x^\mu}{{\rm d}\tau}$ with only $t$ and $\rho$ components, one has from its normalization
\[\left(\frac{{\rm d}\rho}{{\rm d}\tau}\right)^{\!2}
  =\frac{1}{g_{\rho\rho}}\left(\frac{{\cal E}^2}{-g_{tt}}-1\right)
  =\frac{{\cal E}^2-N^2}{e^{2\lambda}} \;,\]
where ${\cal E}\!:=\!-u_t$ is the conserved energy per unit rest mass. Evaluating this at the starting point $(\rho\!=\!0,z\!=\!0)$ where $u^\mu\!=\!N^{-1}\delta^\mu_t$, we have
\[{\cal E}=-g_{tt}u^t=N^2 u^t=(N^2 u^t)_{\rho=0,z=0}=N(\rho\!=\!0,z\!=\!0)\]
and thus
\begin{equation}  \label{flight,timelike}
  \frac{{\rm d}\tau}{{\rm d}\rho}
  =\frac{e^{\lambda(z=0)}}{\sqrt{N^2(z\!=\!0,\rho\!=\!0)-N^2(z\!=\!0)}} \,,
\end{equation}
The time of flight is obtained by integrating this between the desired values of $\rho$. 
For the MP ring, $\lambda\!=\!0$ and by substituting for $N$ one finds that the integrand of (\ref{flight,timelike}) is everywhere finite (except for the equilibrium starting point $\rho\!=\!0$, of course) and going to the value $(a+M)/a$ at the very ring. Hence, the test particle reaches the ring in finite proper time.
For the BW ring, $\lambda$ is given by (\ref{lambda-BW}) and after substitution for $N$ one finds that the integrand of (\ref{flight,timelike}) behaves quite differently at the ring, namely it diverges there as
\[\exp\frac{M}{a}\cdot\,\exp\frac{M^2}{\pi^2 a\,(a-\rho)} \,.\]
Therefore, the proper time necessary to reach the ring is exponentially divergent. This conclusion was already made by \cite{D'AfonsecaLO-05} who observed, by numerical integration and for generic geodesics, that all the free particles that approach the ring do so from its ``inside'' and only at infinite time.

For photons, we can find the {\em coordinate} (Killing) time $t$ spent in reaching the ring along the $(\rho\!<\!a,z\!=\!0)$ circle. Restricting to purely radial geodesics as above, we have, directly from $0=g_{tt}{\rm d}t^2+g_{\rho\rho}{\rm d}\rho^2$,
\begin{equation}  \label{flight,light}
  \frac{{\rm d}t}{{\rm d}\rho}=\frac{e^{\lambda(z=0)}}{N^2(z\!=\!0)} \,.
\end{equation}
For the MP ring, we thus have -- using the result (\ref{MP-radius}) -- 
\begin{align}
  &\int\limits_0^a
    \left[1+\frac{2MK\!\left(\frac{2\sqrt{a\rho}}{\rho+a}\right)}{\pi\,(\rho+a)}\right]^{\!2}
    {\rm d}\rho =  \nonumber \\
  &= a+\frac{8{\cal G}}{\pi}M+\frac{4M^2}{\pi^2 a}\int\limits_0^1 K^2(v)\,{\rm d}v \,,
\end{align}
where everything is finite, including the last integral
\[\int\limits_0^1 K^2(v)\,{\rm d}v \doteq 3.4987815 \,.\]
For the BW ring, on the contrary, the integrand of (\ref{flight,light}) again diverges at $\rho\!\rightarrow\!a$, this time as
\[\left(1-\frac{\rho}{a}\right)^{\!-\frac{2M}{\pi a}}
  \exp\frac{M^2}{\pi^2 a\,(a-\rho)} \;,\]
so the light-travel time $t$ is infinite.

Both the above results confirm the contrast between the MP and the BW rings.

\subsection{Circular equatorial geodesics}

\begin{figure*}
\includegraphics[width=\textwidth]{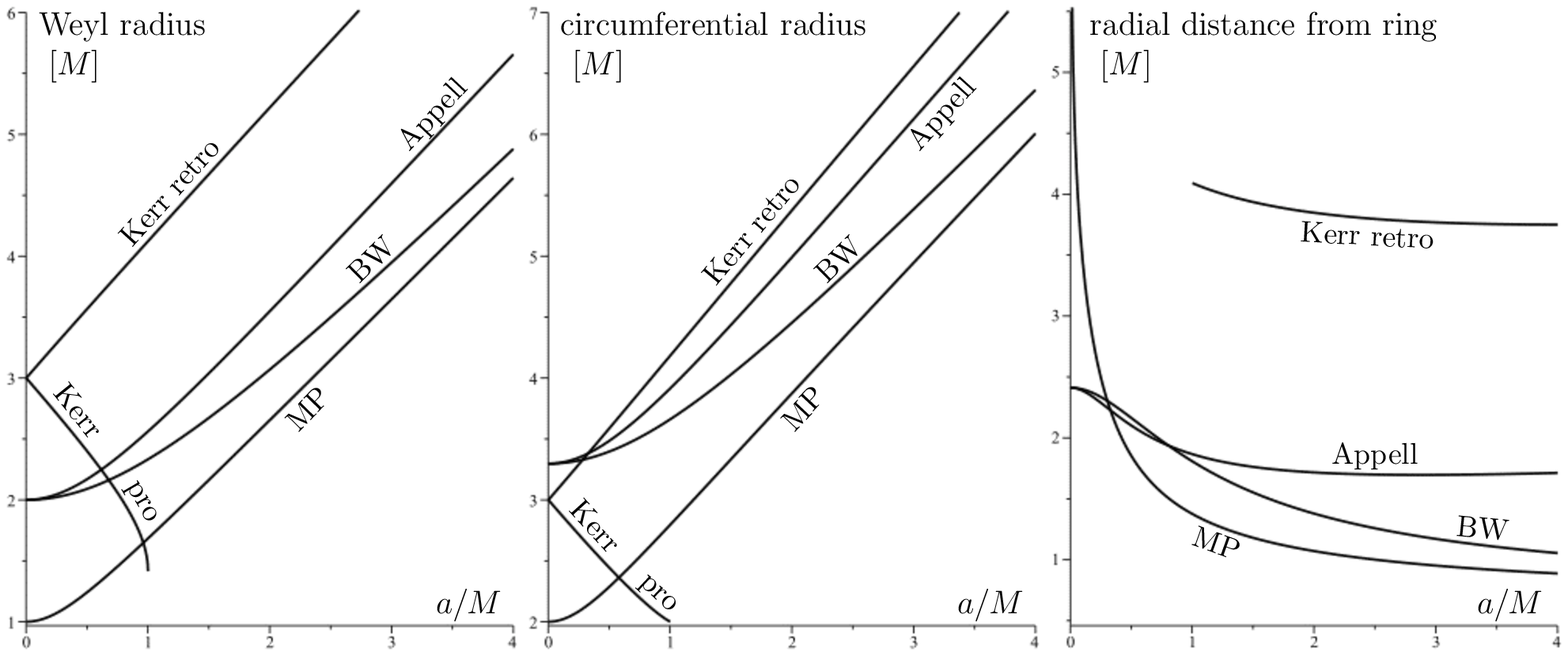}
\caption
{Weyl radius (for Kerr space-time: Kerr-Schild radius) (left plot), circumferential radius (middle plot) and proper radial distance from the ring (right plot) of a photon circular geodesic, plotted in dependence on $a/M$ for the four rings compared in this paper. For Kerr, the black-hole case is limited to $a/M\!\leq\!1$; for $a/M\!>\!1$ (naked-singularity case), the prograde photon orbit lies at $\rho\!=\!a$, so it coincides with the ring itself and is not shown. In Kerr space-time, the proper radial distance from the ring only has a good sense for $a\!>\!M$ case when there is no horizon, so only retrograde orbit is included in the right plot.}
\label{photon-circular}
\end{figure*}

Another simple feature independent of coordinates is the existence of circular geodesics in the ring's plane; in fact even their angular velocity $\Omega\!:=\!{\rm d}\phi/{\rm d}t$ is such, because $t$ and $\phi$ are Killing coordinates and $\Omega$ represents the angular velocity with respect to the asymptotic inertial frame. The most ``absolute'' of course is the photon circular geodesic, and since it is the innermost one, one can also expect that it best reflects the differences between the rings. Let us focus on it.

The equation for photon circular orbit has been derived many times and for the Weyl fields it is very simple,
\begin{equation}
  2\rho\,N_{,\rho}=N \qquad ({\rm i.e.,}\;\;2\rho\,\nu_{,\rho}=1)\,.
\end{equation}
One can loosely say that larger orbital radius indicates stronger source, but it has to be emphasized that $\rho$ is only a coordinate and, mainly, that it is virtually impossible to thus compare {\em different} space-times. Nevertheless, what {\em can} tell something is a corresponding circumferential radius (the one which gives proper circumference of the orbit's spatial track when multiplied by $2\pi$)
\[\sqrt{g_{\phi\phi}(z\!=\!0,\rho\!=\!\rho_{\rm ph})}
  =\frac{\rho_{\rm ph}}{N(z\!=\!0,\rho\!=\!\rho_{\rm ph})}\]
or a {\em proper} radial distance of the orbit from the ring
\[\int\limits_a^{\rho_{\rm ph}}\sqrt{g_{\rho\rho}(z\!=\!0)}\;{\rm d}\rho
  =\int\limits_a^{\rho_{\rm ph}}\frac{e^{\lambda(z=0)}}{N(z\!=\!0)}\;{\rm d}\rho\]
(all the rings are in a finite proper distance from the outer equatorial plane, so this really has a good meaning).

In figure \ref{photon-circular} we show, for our four rings, how the photon-orbit Weyl (or Kerr-Schild) radius, circumferential radius and proper radial distance from the ring depend on the ring radius $a$ (for a unit ring mass $M$). One sees that the coordinate-radius values provide quite a good picture this time. The proper distance of the photon orbit from the ring (right plot) decreases with $a/M$ for the MP and BW cases, whereas for the Appell and Kerr cases it only decreases until $a$ reaches several $M$ and then increases, on the contrary (this region is not shown already). Another peculiar feature is the (logarithmic) divergence of the equatorial radial distance from the MP ring at $a\!\rightarrow\!0^+$ (from other rings it remains finite in this limit); however, this is consistent with the fact that the MP ring becomes an extreme horizon in that limit. We add that there are no circular orbits in the $R\!<\!0$ sheet of the Appell and Kerr space-times, since there the field is repulsive ($M$ acts with minus sign).

\subsection{Gravitational acceleration}

\begin{figure*}
\includegraphics[width=0.9\textwidth]{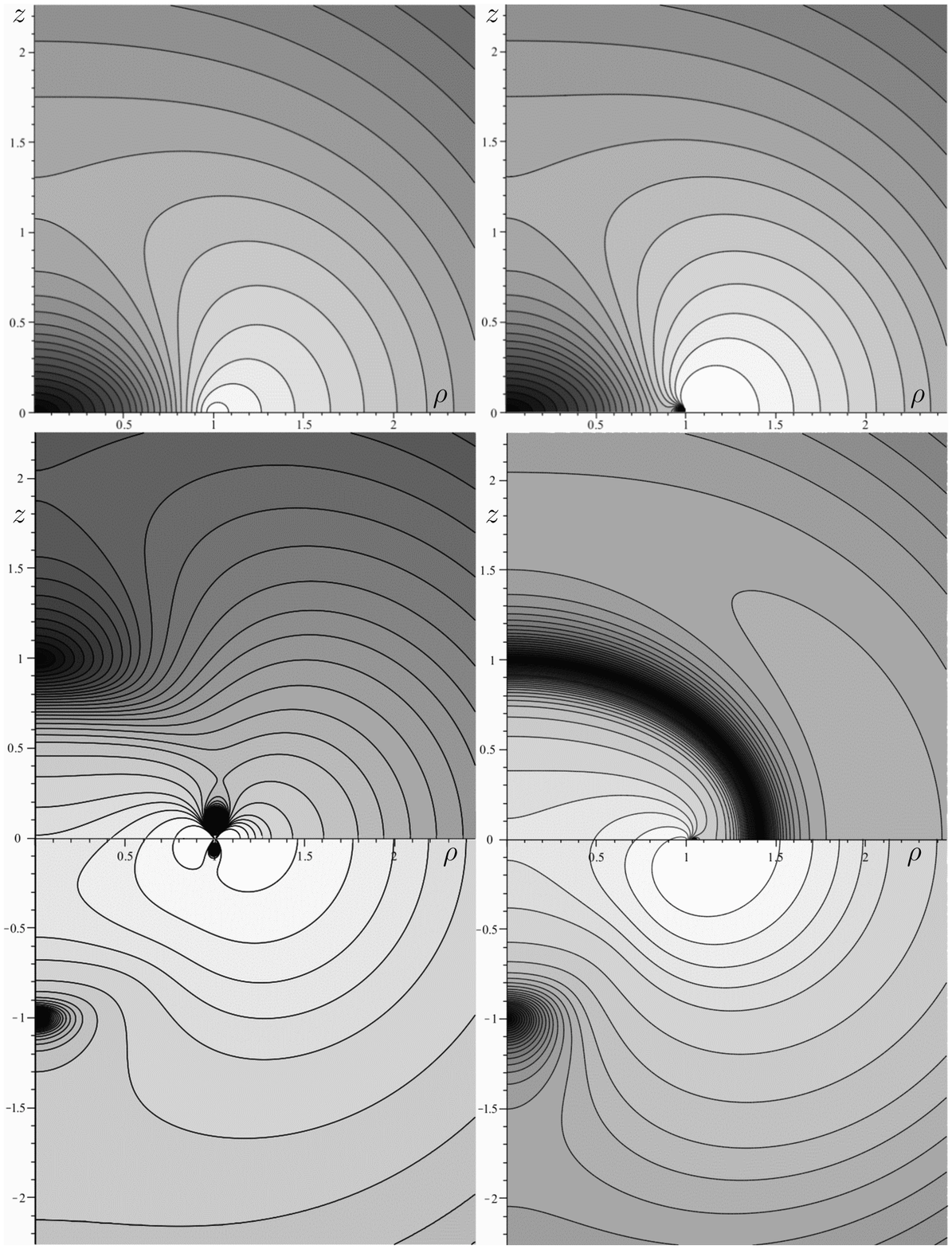}
\caption
{Contours of the gravitational acceleration $\kappa$ given by (\ref{kappa}), plotted within the meridional plane for the MP ring (top left), the BW ring (top right), the Appell ring (bottom left) and the Kerr ring (bottom right). All the rings have $M\!=\!a$. The $(\rho,z)$ axes (scaled by $a$) represent Weyl coordinates for the static rings, while Kerr-Schild coordinates for the Kerr ring. We take $\kappa$ as positive everywhere (as {\em plus} square root of $\kappa^2$), with light/dark shading indicating larger/smaller values. The contour levels are same for all the rings, in the units of $1/a^2$ they range from $0$ to $0.25$, only close to the Appell and Kerr rings they extend up to $15$. The (extreme) Kerr horizon is visible as the black half-arc where $\kappa\!=\!0$.}
\label{acceleration}
\end{figure*}

The gradient of $N$ provides an important invariant quantity, the ``gravitational acceleration'' (field intensity) $\kappa$, given by
\begin{equation}  \label{kappa}
  \kappa^2:=g^{\mu\nu}N_{,\mu}N_{,\nu}
  =\frac{N^2}{e^{2\lambda}}\left[(N_{,\rho})^2+(N_{,z})^2\right].
\end{equation}
This scalar is mainly given by the ``Newtonian'' quantity $N$, only containing $\lambda$ because of calculating the {\em magnitude} of the gradient. But there is no analogous invariant given by $\lambda$, simply because $g_{\rho\rho}\!=\!g_{zz}$ has no invariant meaning like $g_{tt}$ does have due to the Killing time symmetry.
Figure \ref{acceleration} illustrates how the acceleration appears for our rings with $M\!=\!a$. In looking at it, one may recall for comparison that on stationary (Killing) horizons, this quantity (surface gravity) is constant.

For the MP ring, for which $\lambda\!=\!0$, the acceleration is zero at the very centre ($\rho\!=\!0$, $z\!=\!0$) and also vanishes toward spatial infinity, with maximum around the ring radius; this maximum is divergent in the ring plane, whereas it is finite elsewhere. On the symmetry axis, in particular,
\[\kappa^2(\rho\!=\!0)=\frac{M^2 z^2}{\left(M+\sqrt{z^2+a^2}\right)^6} \,.\]
One thus observes again that the ring does not correspond to an extreme horizon, because on horizons $\kappa$ is finite (being called surface gravity there), especially on extreme ones it even vanishes.

For the Bach-Weyl ring, the acceleration behaves similarly, reducing to just
\[\kappa^2(\rho\!=\!0)
  =\frac{M^2 z^2}{(z^2+a^2)^3\,\exp\frac{4M}{\sqrt{z^2+a^2}}}\]
on the symmetry axis (and hence vanishing at the very central point).

In the Appell-ring field, the acceleration expression is slightly more complicated, but on the axis it reduces to
\[\kappa^2(\vartheta\!=\!0)
  =\frac{M^2}{\exp\frac{4MR}{R^2+a^2}}\;\frac{(R^2-a^2)^2}{(R^2+a^2)^4} \;,\]
specially at $\vartheta\!=\!0$, $R\!=\!0$ one has $\kappa\!=\!M^2/a^4$ (pointing from the $R\!<\!0$ to the $R\!>\!0$ space). Obviously there are two zero-valued local extrema of $\kappa^2$ on the axis, at $R\!=\!\pm a$.
The acceleration scalar also simplifies considerably on the disc inside the ring,
\[\kappa^2(R\!=\!0)=\frac{M^2}{a^4\cos^6\vartheta}\,\exp\frac{M^2(1-\cos^4\vartheta)}{2a^2\cos^4\vartheta} \;,\]
as well as in the equatorial plane,
\[\kappa^2(\vartheta\!=\!\pi/2)=\frac{M^2(R^2+a^2)}{R^6\,\exp\frac{M\,[2R^2(4R-M)-Ma^2]}{2R^4}} \;.\]
It is also worth to write down the acceleration square on the boundary $|R|\!=\!a\cos\vartheta$ of the ``repulsive'' region,
\[\kappa^2(R\!=\!\pm a\cos\vartheta)
  =\frac{M^2\sin^2\vartheta}{8a^4\cos^6\vartheta}\,
   \frac{1}{\exp\frac{M(M\sin^4\vartheta\pm 16a\cos^3\vartheta)}{8a^2\cos^4\vartheta}} \;.\]
Finally, in the $a\!\rightarrow\!0^+$ limit the result is the same as in the BW-ring case, reading
\[\kappa^2(a\!\rightarrow\!0^+)=\frac{M^2}{R^4\,\exp\frac{M(4R-M\sin^2\vartheta)}{R^2}} \;;\]
this is finite except at the $R\!=\!0$ central circle (which is just a point in that limit, however).

In the Kerr field, the acceleration square reads\footnote
{Physically, it represents square of the acceleration of the stationary circular motion with zero angular momentum (the well known ZAMO observers), reparametrized with respect to the asymptotic inertial time $t$.}
\[\kappa^2=M^2\,
           \frac{\Sigma^2(r^4\!-\!a^4)^2+4r^4\Delta\left[2\Sigma(r^2\!+\!a^2)\!-\!{\cal A}\right]a^2\sin^2\theta}
                {\Sigma^2\,{\cal A}^3} \,,\]
with special values
\begin{align*}
  \kappa^2(\theta\!=\!0)     &= \frac{M^2(r^2-a^2)^2}{(r^2+a^2)^4} \;, \\
  \kappa^2(\theta\!=\!\pi/2) &= \frac{M^2[(r^2+a^2)^2-4Mra^2]^2}{r^3(r^3+ra^2+2Ma^2)^3} \;, \\
  \kappa^2(r\!=\!0) &= \frac{M^2}{a^4\cos^6\theta} \;, \\
  \kappa^2(a\!=\!0) &= \frac{M^2}{r^4} \;.
\end{align*}
On the horizon (given by $N\!=\!0$ $\Leftrightarrow$ $\Delta\!=\!0$, the larger root of which is $r\!=\!r_{\rm H}\!:=\!M\!+\!\sqrt{M^2-a^2}$), the scalar is called {\it surface gravity} and gets very simple (most importantly, it is independent of $\theta$),
\[\kappa^2(r\!=\!r_{\rm H})=\frac{M^2-a^2}{(r_{\rm H}^2+a^2)^2}
                           =\frac{M^2-a^2}{(2Mr_{\rm H})^2} \;;\]
in particular, it vanishes if the horizon is extreme ($M\!=\!a$).
Interestingly, $\kappa^2$ diverges not only at the singularity ($\Sigma\!=\!0$), but also where ${\cal A}\!=\!0$, i.e. on the boundary of the chronology-violating region existing in the $r\!<\!0$ half-space.

Again, the rings differ strongly in the behaviour of the acceleration in their closest vicinity: in the ring-focused toroidal coordinates $(\zeta,\psi)$, in which the ring is approached as $\zeta\!\rightarrow\!\infty$, we have in this limit
\begin{align*}
  \kappa^2 &\sim \frac{\pi^4 a^2\,e^{2\zeta}}{M^4\,\zeta^4}
           &\,& \dots\;{\rm MP~ring} \,, \\
           &\sim \frac{M^2\zeta^2}{\pi^2 a^4}\;\exp\left(\frac{4M^2\zeta^2}{\pi^2 a^2}\;e^\zeta\cos\psi\right)
           &\,& \dots\;{\rm BW~ring} \,, \\
           &\sim \frac{M^2}{64a^4}\;\exp\left(\frac{M^2}{32a^2}\;e^{2\zeta}\cos 2\psi\right)
           &\,& \dots\;{\rm Appell~ring} \,, \\
           &\sim \frac{1}{64Ma}\;\frac{\exp\frac{3\zeta}{2}}{\cos^3\frac{\psi}{2}}
           &\,& \dots\;{\rm Kerr~ring} \,.
\end{align*}
Hence, the MP ring is the only which -- in this respect and in this coordinates -- is isotropic.

\section{Curvature}
\label{curvature}

Now we proceed to the level of the second metric derivatives, hence to the level of field equations. Let us just briefly recall that for any static axisymmetric (electro)vacuum, i.e. for the energy-momentum tensor
\begin{equation}  \label{Tmunu,EM}
  T_{\mu\nu}=\frac{1}{4\pi}\left(F_{\mu\lambda}{F_\nu}^\lambda-
                                 \frac{1}{4}\,g_{\mu\nu}F_{\kappa\lambda}F^{\kappa\lambda}\right),
\end{equation}
with $F_{\mu\nu}\!\equiv\!A_{\nu,\mu}\!-\!A_{\mu,\nu}$ denoting the electromagnetic-field tensor and $A_\mu$ the electromagnetic four-potential, and with zero cosmological constant, the field equations reduce to (e.g. \cite{Carminati-81})
\begin{align}
  \nu_{,\rho\rho}\!+\frac{\nu_{,\rho}}{\rho}+\nu_{,zz}
    &= 4\pi\,\frac{e^{2\lambda}}{N^2}\,(T^\phi_\phi-T^t_t)  \nonumber \\
    &= \frac{e^{2\lambda}}{N^2}\,(F_{\phi\lambda}F^{\phi\lambda}\!-\!F_{t\lambda}F^{t\lambda})  \nonumber \\
    &= \frac{1}{N^2}\left[(\Phi_{,\rho})^2\!+\!(\Phi_{,z})^2\right], \label{Laplace,nu,elstat} \\
  \lambda_{,\rho}\!-\rho(\nu_{,\rho})^2\!+\rho(\nu_{,z})^2
    &= 4\pi\rho\,(T_{\rho\rho}-T_{zz})  \nonumber \\
    &= \rho\,(F_{\rho\lambda}{F_\rho}^\lambda\!-\!F_{z\lambda}{F_z}^\lambda)  \nonumber \\
    &= {-}\frac{\rho}{N^2}\left[(\Phi_{,\rho})^2\!-\!(\Phi_{,z})^2\right], \label{Weyl-lambda,rho,elstat} \\
  \lambda_{,z}\!-2\rho\,\nu_{,\rho}\nu_{,z}
    &= 8\pi\rho\,T_{\rho z}  \nonumber \\
    &= 2\rho\,F_{\rho\lambda}{F_z}^\lambda  \nonumber \\
    &= {-}\frac{2\rho}{N^2}\;\Phi_{,\rho}\Phi_{,z} \,, \label{Weyl-lambda,z,elstat}
\end{align}
where the right-hand sides' third forms are obtained by introducing a scalar potential $\Phi(\rho,z)$ as
\begin{equation}
  A_\mu\!=\!(-\Phi,0,0,0)
  \quad \Longrightarrow \;\;
  F_{t\rho}=\Phi_{,\rho} \,, \; F_{tz}=\Phi_{,z} \,.
\end{equation}
Besides these three, one has to also consider the Maxwell equations which in the electrostatic case have only one non-trivial component
\begin{equation}
  \Phi_{,\rho\rho}+\frac{\Phi_{,\rho}}{\rho}+\Phi_{,zz}
    =2\nu_{,\rho}\Phi_{,\rho}+2\nu_{,z}\Phi_{,z} \,.
\end{equation}

Since the energy-momentum tensor is traceless, $T^\nu_\nu\!=\!0$, the Einstein equations (without cosmological term) imply that the Ricci scalar $R^\nu_\nu$ is zero as well, which yields useful relation
\begin{align}
  & N_{,\rho\rho}+\frac{N_{,\rho}}{\rho}+N_{,zz} \;(=:\!\Delta N)\, = \nonumber \\
  &= {} \frac{2}{N}\left[(N_{,\rho})^2+(N_{,z})^2\right]+
     {} N(\lambda_{,\rho\rho}+\lambda_{,zz}).  \label{R=0}
\end{align}
Of the two non-trivial and independent invariants of the electromagnetic-field tensor, $F_{\mu\nu}F^{\mu\nu}$ and $F_{\mu\nu}{^*\!F}^{\mu\nu}$, the second (given by the dual tensor ${^*\!F}^{\mu\nu}$) vanishes in the static situation.
It is also easy to find that
\begin{align}
  R_{\mu\nu}R^{\mu\nu} &=(F_{\mu\nu}F^{\mu\nu})^2, \\
  F_{\mu\nu}F^{\mu\nu} &=-2e^{-2\lambda}\left[(\Phi_{,\rho})^2+(\Phi_{,z})^2\right],
\end{align}
as well as to check that
$\,4\pi(T^\rho_\rho\!+\!T^z_z)=F^{\rho z}F_{\rho z}\!-\!F^{t\phi}F_{t\phi}$
vanishes, as required for the Weyl form of the metric (\ref{metric-Weyl}).

For the MP ring, the above relations simplify due to the relations $\Phi\!=\!\mp N$ and (consequently) $\lambda\!=\!0$, valid for all Majumdar-Papapetrou solutions, while the BW and Appell rings are vacuum, so $T_{\mu\nu}\!=\!0$ in their case.

Before proceeding to curvature invariants of the four-dimensional space-time -- the Kretschmann scalar, in particular --, let us turn to a two-dimensional characteristic, the Gauss curvature, which often provides a better insight. Actually, since the Kretschmann scalar (as well as other 4D invariants) also contains a ``time contribution'' which has no immediate tidal meaning, it is not necessarily the most intuitive curvature characteristic, while the Gauss curvature of a suitably chosen 2D surface can reveal the spatial geometry better. In the Weyl-metric case, the most privileged surfaces are the meridional planes $\{t\!=\!{\rm const},\phi\!=\!{\rm const}\}$ and the equatorial plane $\{t\!=\!{\rm const},z\!=\!0\}$.

\subsection{Gauss curvature of the meridional plane}

\begin{figure*}
\includegraphics[width=0.9\textwidth]{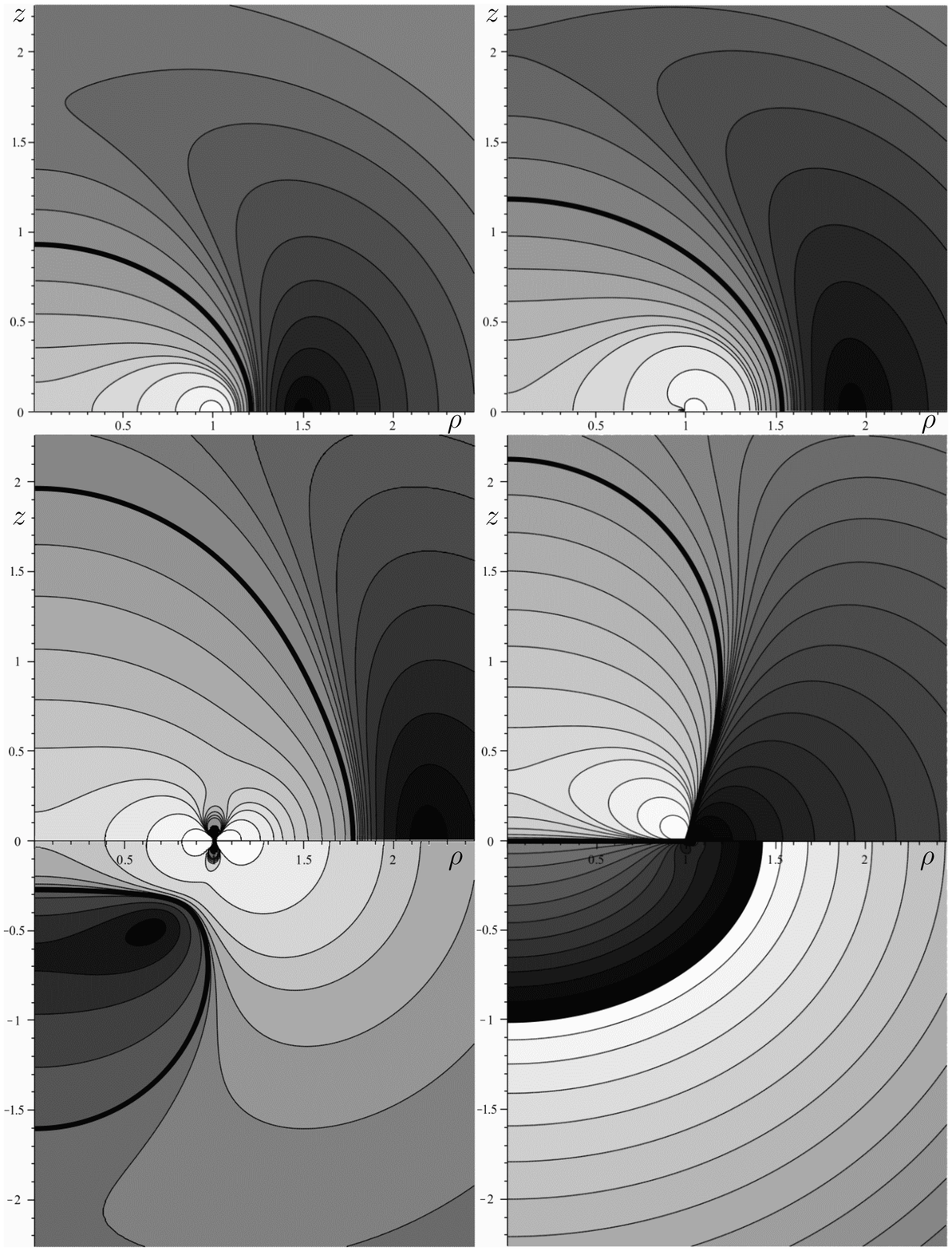}
\caption
{Contours of the Gauss curvature (\ref{Gauss,phi=const}) of the $\phi\!=\!{\rm const}$ surfaces (meridional planes), shown for the Majumdar-Papapetrou ring (top left), the Bach-Weyl ring (top right), the Appell ring (bottom left) and the Kerr ring (bottom right). For the Kerr ring, the Kerr-Schild plane $\{T\!=\!{\rm const},\psi\!=\!{\rm const}\}$ is rather taken as ``meridional'' -- see the main text. All the rings again have $M\!=\!a$. Light/dark shading indicates positive/negative values, zero-value contour is emphasized. In the units of $1/a^2$, the contour levels range from $-0.034$ to $0.6$ for the MP ring, from $-0.027$ to $8$ for the BW ring, from $-1.5$ to $2000$ for the Appell ring, and from $-4$ to $+3$ ($-1000$ to $+800$ for the ``bottom" sheet) for the Kerr ring.}
\label{Gauss-phi}
\end{figure*}

\begin{figure*}
\includegraphics[width=0.9\textwidth]{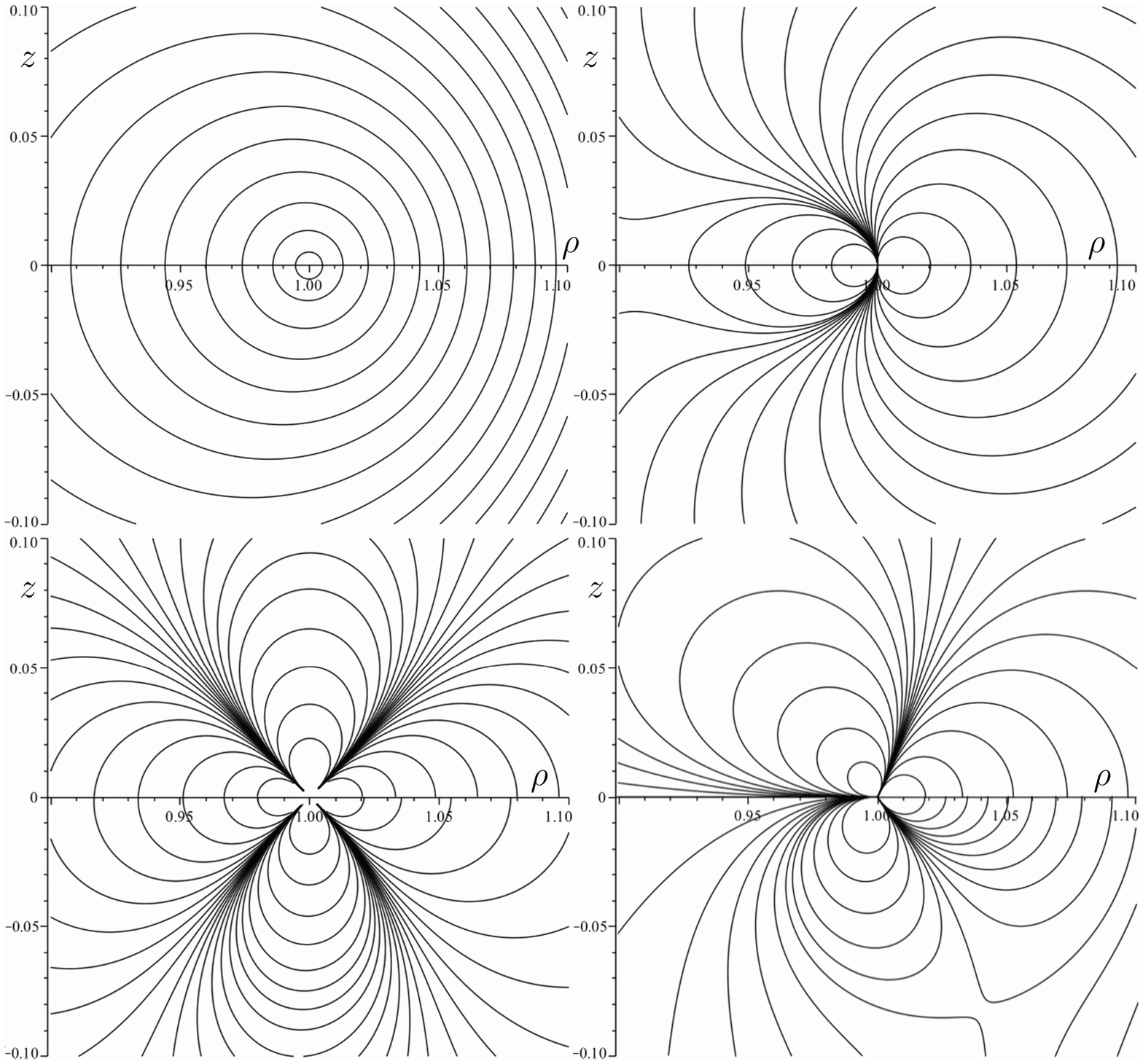}
\caption
{Details of plots shown in figure \ref{Gauss-phi}, magnifying the region in a close vicinity of the ring sources ($0.9\!<\!\rho/a\!<\!1.1$, $-0.1\!<\!z/a\!<\!0.1$). Again the Majumdar-Papapetrou ring is at top left, the Bach-Weyl ring at top right, the Appell ring at bottom left and the Kerr ring at bottom right. The rings are clearly very different, with only the Majumdar-Papapetrou ring providing a satisfactory (locally cylindrical) picture. In the units of $1/a^2$, the contour levels now range from $0.1$ to $30$ for the MP ring, from $0.001$ to $1.35\cdot 10^6$ for the BW ring, from $10^{-100}$ to $10^{140}$ for the Appell ring, and from $-400$ to $+400$ for the Kerr ring.}
\label{Gauss-phi-detail}
\end{figure*}

From the meridional restriction of the metric and by employing (\ref{R=0}), one obtains the Gauss curvature (half of the corresponding 2D Ricci scalar)
\begin{equation}  \label{Gauss,phi=const}
  \frac{^{(2)\!}R}{2}=\left[(N_{,\rho})^2-\frac{NN_{,\rho}}{\rho}+(N_{,z})^2\right]e^{-2\lambda} \,.
\end{equation}
For the MP ring, this yields quite a long expression which reduces to
\[\frac{{^{(2)}\!}R\,(\rho\!=\!0,z\!=\!0)}{2}=\frac{M}{2\,(M+a)^3}\]
at the very centre. For the BW ring, the central value is
\[\frac{{^{(2)}\!}R\,(\rho\!=\!0,z\!=\!0)}{2}=\frac{M}{2a^3\exp\frac{2M}{a}} \;.\]
For the Appell ring, the axial and equatorial behaviours are tractable (and slightly shorter in the oblate coordinates $R$, $\vartheta$ -- see (\ref{oblate})),
\begin{align*}
 &\frac{{^{(2)}\!}R\,(\vartheta\!=\!0)}{2}
     = \frac{M^2(R^2\!-\!a^2)^2\!-MR\,(R^2\!-\!3a^2)(R^2\!+\!a^2)}
            {(R^2+a^2)^4\;\exp\frac{2\,MR}{R^2+a^2}} \,, \\
 &\frac{{^{(2)}\!}R\,(\vartheta\!=\!\pi/2)}{2}
     = -\frac{M\,[R^3-M(R^2+a^2)]}{R^6\,\exp\frac{M\,(4R^3-2MR^2-Ma^2)}{2R^4}} \;, \\
 &\frac{{^{(2)}\!}R\,(R\!=\!0)}{2}
     = \frac{M^2}{a^4\cos^6\vartheta}\;\exp\frac{M^2(1\!+\!\cos^2\vartheta)\sin^2\vartheta}{2a^2\cos^4\vartheta} \;,
\end{align*}
and reducing to just
\[\frac{{^{(2)}\!}R\,(R\!=\!0,\vartheta\!=\!0)}{2}=\frac{M^2}{a^4}\]
at the central point.

The results obtained for all the rings are illustrated in figures \ref{Gauss-phi} (the whole central region of the meridional plane) and \ref{Gauss-phi-detail} (zoom to the closest coordinate vicinity of the rings), including also the Gauss curvature of the Kerr meridional plane. The latter needs a commentary.
In {\em static} axisymmetric space-times, it is pretty clear what to understand by ``meridional planes'': they are simply orthogonal to both the existing Killing symmetries. In the {\em rotating} (generic stationary) case, however, it is not that clear, because of the presence of differential dragging in the azimuthal direction. Dragging means different angular velocity of a ``physical meridional plane'' at different radii (and also latitudes), so the plane stretches necessarily along azimuthal direction and is wound about the centre. At the Kerr black-hole horizon, the angular velocity in the Boyer-Lindquist coordinates, ${\rm d}\phi/{\rm d}t$, of any free test particle remains finite, but both ${\rm d}\phi$ and ${\rm d}t$ diverge there. Hence, it is desirable to turn to different coordinates where time and azimuth would behave better and then fix the meridional plane with respect to them. Such a more reasonable choice is just provided by the Kerr-Schild coordinates (\ref{Kerr-Schild}).
The plane $\{T\!=\!{\rm const},\psi\!=\!{\rm const}\}$ is described, as expressed in the Boyer-Lindquist coordinates, by the metric
\[{\rm d}l^2=\frac{\Sigma}{r^2+a^2}\left(1+\frac{2Mr}{r^2+a^2}\right){\rm d}r^2+\Sigma\,{\rm d}\theta^2 \,,\]
and its Gauss curvature reads
\[-Mr\;\frac{(r^2\!+\!a^2)^2(r^2\!-\!3a^2\cos^2\theta)-4Mra^2(\Sigma\!-\!2r^2\sin^2\theta)}
        {\Sigma^3\,(r^2+2Mr+a^2)^2} \,.\]
This vanishes on $r\!=\!0$, while on the axis and in the equatorial plane it reduces to
\begin{align*}
  {\rm axis:}  &\quad -Mr\,\frac{r^4-a^4-2a^2(r^2+2Mr+a^2)}{(r^2+a^2)^2(r^2+2Mr+a^2)^2} \;, \\
  {\rm equat:} &\quad -\frac{M}{r^3}\,\frac{r^4-a^4+2a^2(r^2+2Mr+a^2)}{(r^2+2Mr+a^2)^2} \;.
\end{align*}
In the $r\!>\!0$ half-space, the Gauss curvature behaves similarly as around static rings, except for that the zero-value contour ends at the (Kerr) ring in contrast to the static-ring cases. In the $r\!<\!0$ half-space, the picture is very different (even from the Appell-ring case which is itself interesting), with a kind of ``anti-horizon'' appearing at $r^2\!+2Mr+a^2\!=\!0$ where the curvature jumps (in the direction of decreasing $r$) from negative to positive infinity.

Checking the behaviour at the very rings (i.e. in the $\zeta\!\rightarrow\!\infty$ limit in toroidal coordinates), one finds
\begin{align*}
  \frac{^{(2)\!}R}{2}
      &\sim \frac{\pi^2\,e^{2\zeta}}{M^2\,\zeta^2}
      &\,& \dots\;{\rm MP~ring} \,, \\
      &\sim \frac{M^2\zeta^2}{\pi^2 a^4}\;\exp\left(\frac{4M^2\zeta^2}{\pi^2 a^2}\;e^\zeta\cos\psi\right)
      &\,& \dots\;{\rm BW~ring} \,, \\
      &\sim \frac{M^2}{64a^4}\;\exp\left(\frac{M^2}{32a^2}\;e^{2\zeta}\cos 2\psi\right)
      &\,& \dots\;{\rm Appell~ring} \,, \\
      &\sim \frac{M}{8a^3}\;e^{3\zeta/2}\,(1\!-\!2\cos\psi)\cos\frac{\psi}{2}
      &\,& \dots\;{\rm Kerr~ring} \,.
\end{align*}
Again the MP ring is the only isotropic case.

\subsection{Gauss curvature of the equatorial plane}

\begin{figure*}
\includegraphics[width=\textwidth]{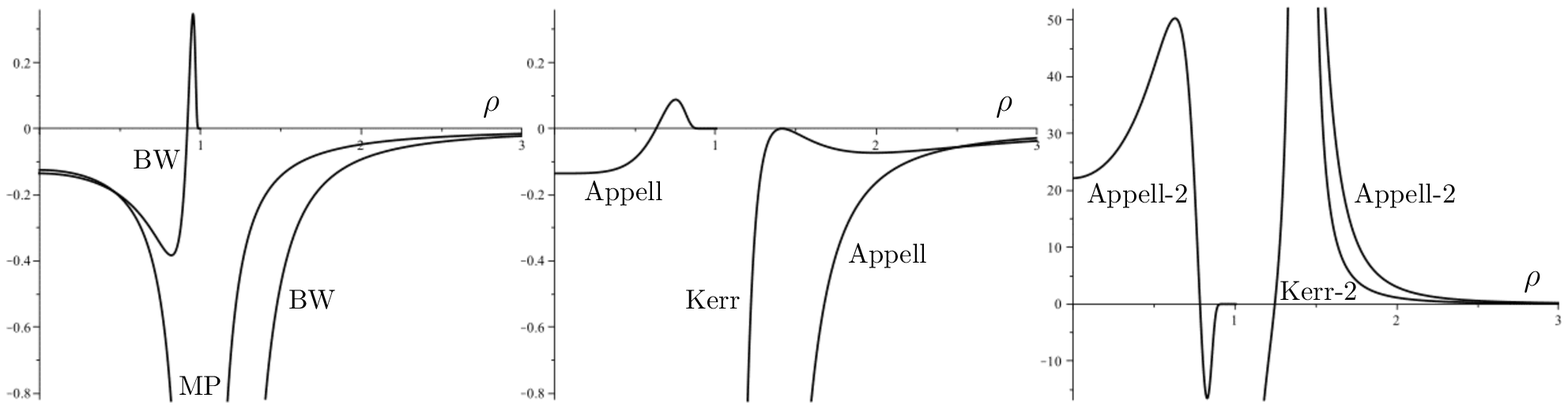}
\caption
{Gauss curvature (\ref{Gauss,z=0}) of the equatorial plane, plotted against $\rho$ for all the four rings, setting $M\!=\!a$; $\rho$ again represents Weyl radius for the three static rings, while Kerr-Schild radius in the Kerr case. The two left plots are in exactly the same scale, while the right-hand one covers much wider vertical range. Starting from the left plot, the curvature tends to negative values and diverges at the very rings, with the MP ring generating a ``natural'' behaviour, while the BW curve being much more wild {\em below} the ring; the Appell-ring curve is qualitatively similar to the latter, but much smoother; the Kerr ring generates {\em zero} curvature on the enclosed circle, while outside there exists a maximum with exactly zero value (this however is specific for the extreme, $M\!=\!a$ case: for $M\!>\!a$ the maximum would be positive and for $M\!<\!a$ the whole curve would be negative). The rightmost plot shows how wild is the situation on the second, negative-mass sheets of the Appell and Kerr solutions (note again the much wider range of the vertical axis); the Gauss curvature tends to {\em positive} values there, with the most peculiar feature being the (positive) divergence occurring for Kerr at the boundary ${\cal A}\!=\!0$ of the chronology-violating region.}
\label{Gauss-equat}
\end{figure*}

Another privileged section is the equatorial plane, $\{t\!=\!{\rm const},z\!=\!0\}$, the more that the fields of all the rings considered in this paper are reflection symmetric with respect to it. Restricting the metric to this plane and computing its 2D Ricci scalar, one finds that its Gauss curvature is
\begin{equation}  \label{Gauss,z=0}
  \frac{^{(2)\!}R}{2}=\!\left[NN_{,\rho\rho}\!-\!(N_{,\rho})^2\!+\frac{NN_{,\rho}}{\rho}
                            -\!N\lambda_{,\rho}\!\left(\!N_{,\rho}\!-\!\frac{N}{\rho}\right)\!\right]
                      e^{-2\lambda} .
\end{equation}
Radial behaviour of this quantity in the central region is shown in figure \ref{Gauss-equat} for all four rings again (including the $r\!<\!0$ parts of the Appell and Kerr space-times); differences between the rings are quite big obviously. Leaving aside the $r\!<\!0$ parts of the Appell and Kerr space-times where the equatorial curvature behaves quite wildly (in the Appell case, this essentially looks like reversal of the $r\!>\!0$ behaviour, but within much wider range), the values at the $\rho\!=\!0$, $z\!=\!0$ centre are almost the same for all the static rings, but on the way from the centre towards the rings, the curvature changes in a rather dissimilar manner; in particular, the MP-ring--generated curvature behaves differently from that plotted for the other rings, namely, as the MP ring is approached ``from inside'', the curvature falls to negative infinity monotonically, whereas for the BW and Appell rings it turns positive and finally vanishes at the very ring. (For Kerr, the Gauss curvature of the central circle is zero everywhere.) For the MP and BW rings, interestingly, the Gauss curvature of the equatorial plane at the central point $\rho\!=\!0$, $z\!=\!0$ is exactly {\em minus twice} the Gauss curvature of the meridional planes there, namely it amounts to
\begin{align*}
  \frac{^{(2)\!}R(\rho\!=\!0,z\!=\!0)}{2}
      &= -\frac{M}{(M+a)^3}                   &\,& \dots\;{\rm MP~ring} \,, \\
      &= -\frac{M}{a^3\exp\frac{2M}{a}}       &\,& \dots\;{\rm BW~ring} \,, \\
      &= -\frac{M(2a-M)}{a^4\exp\frac{2M}{a}} &\,& \dots\;{\rm Appell~ring} \,.
\end{align*}

Finally, the Kerr equatorial plane ($\theta\!=\!\pi/2$) has Gauss curvature
\[M\frac{Ma^2(11r^4\!+\!2a^2 r^2\!+\!8Mra^2\!-\!5a^4)\!-\!r(r^2\!+\!3a^2)(r^2\!+\!a^2)^2}
           {r^4\,(r^3+ra^2+2Ma^2)^2}\]
while the Kerr central circle ($r\!=\!0$) is flat (its Gauss curvature is zero). The denominator in the above expression is given by
\[{\cal A}(\theta\!=\!\pi/2)=r\,(r^3+ra^2+2Ma^2)\]
which vanishes at the boundary of the chronology-violating region, namely at
\[\frac{r}{M}=\frac{a^{2/3}}{M^{2/3}}\left(\!\sqrt{1\!+\!\frac{a^2}{27M^2}}-\!1\!\right)^{\!1/3} \!\!\!
              -\frac{\frac{1}{3}\,\frac{a^{4/3}}{M^{4/3}}}
                    {\left(\!\sqrt{1\!+\!\frac{a^2}{27M^2}}-\!1\!\right)^{\!1/3}} \;,\]
so the equatorial Gauss curvature is infinite there.

\subsection{Gauss curvature of the $\rho\!=\!a$ ``cylinder''}

\begin{figure}
\includegraphics[width=\columnwidth]{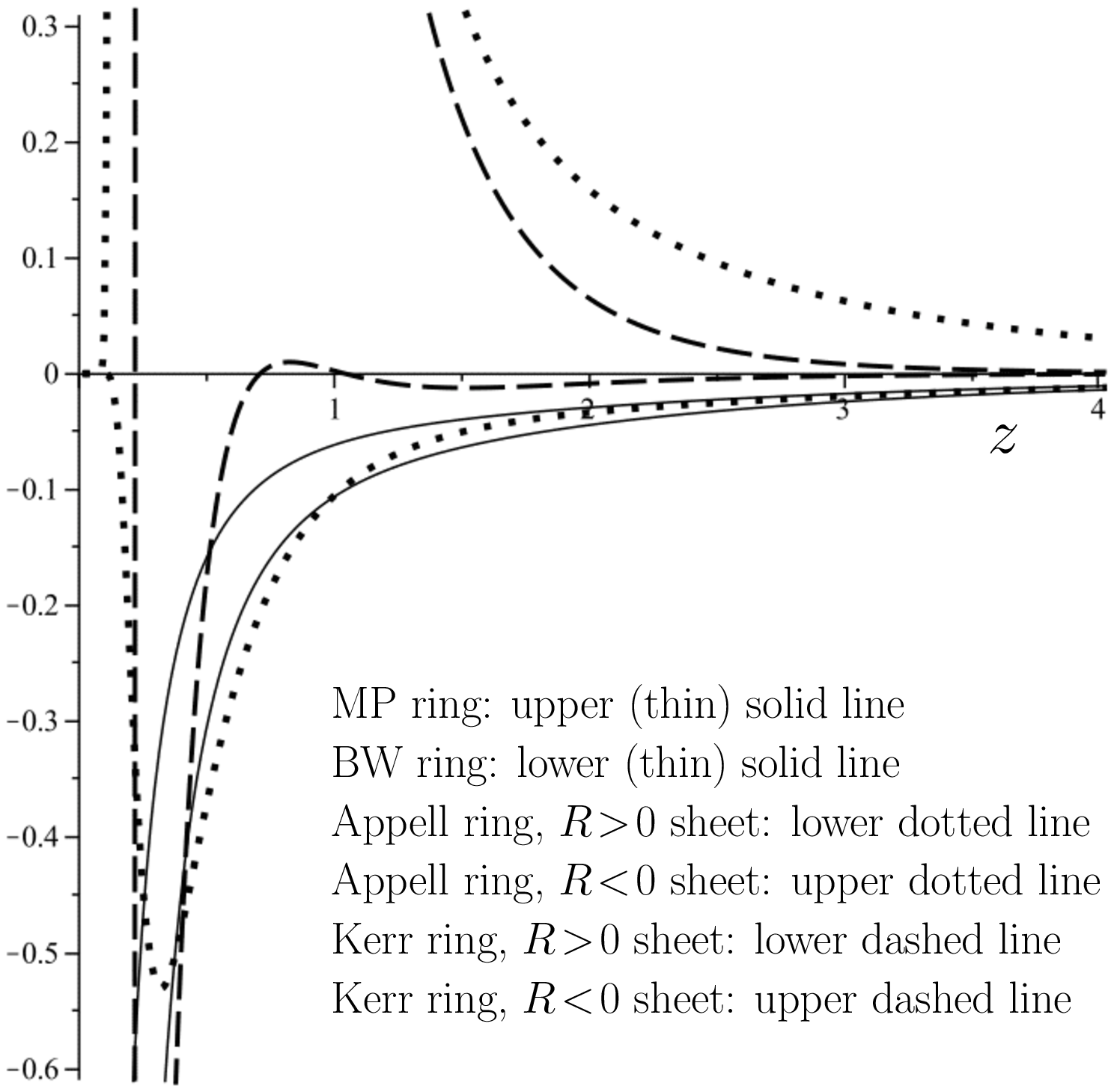}
\caption
{Gauss curvature (\ref{Gauss,rho=a}) of the ``cylinder'' $\rho\!=\!a$, plotted against $z$ (up to $z\!=\!5M$) for all the four rings, while setting $M\!=\!a$; we again denote by $\rho$ and $z$ the Weyl coordinates for the three static rings, while the Kerr-Schild coordinates in the Kerr case. For the MP and BW rings (upper/lower thin solid line) the curvature simply increases monotonically (and finally diverges) when the ring is approached from infinity. For the Appell ring (dotted lines), the curvature is negative/positive in the $R\!>\!0$/$R\!<\!0$ sheet; it again increases (in magnitude) when approaching from a remote region, but, interestingly, after reaching a minimum/maximum (the maximum has much larger magnitude) at certain (different) $z\!<\!a$, it returns to zero at the very ring. In the Kerr-case (dashed line) the Gauss curvature has to be computed by a different formula (see the main text); in the $R\!>\!0$ sheet, the result is similar to that obtained for the MP and BW rings, but before falling to $-\infty$ the curve goes through zero slightly above $z\!=\!a$ and has a low maximum below that value; in the $R\!<\!0$ sheet, the curvature diverges at a certain non-zero $z\!<\!a$ and then, for still lower $z$, it falls from positive to negative infinity extremely steeply.}
\label{Gauss-rho}
\end{figure}

Within the $\{t\!=\!{\rm const}\}$ spaces, the last foliation -- complementary to those by the meridional planes $\phi\!=\!{\rm const}$ and by the horizontal sections $z\!=\!{\rm const}$ -- is represented by ``cylinders'' $\rho\!=\!{\rm const}$. Restricting the metric (\ref{metric-Weyl}) to these $(\phi,z)$ surfaces, one finds that their Gauss curvature reads
\begin{equation}  \label{Gauss,rho=a}
  \frac{^{(2)\!}R}{2}=\left[NN_{,zz}-(N_{,z})^2\right]e^{-2\lambda}.
\end{equation}
For the MP ring the exponential term reduces to unity. In order to again compare the results obtained for the three static rings with that for the Kerr ring, we take, for the latter, the $\{t\!=\!{\rm const},\rho\!=\!{\rm const}\}$ surfaces, i.e. those given by constant Kerr-Schild cylindrical radius $\rho=\sqrt{r^2\!+\!a^2}\,\sin\theta$ (and constant Killing time $t$).

In figure \ref{Gauss-rho}, we specifically show, for the $M\!=\!a$ case, the Gauss curvature of the $\rho\!=\!a$ surface, i.e. of the symmetric cylindrical surface which crosses the equatorial plane just at the ring. In the MP, BW and Kerr cases, the curvature diverges to negative infinity at the rings, whereas in the Appell case it vanishes there (after reaching a finite extreme at certain $z\!<\!a$). Both sheets of the Appell as well as Kerr space-times are considered, like in previous sections. Similarly as in figure \ref{Gauss-equat} (Gauss curvature of the equatorial planes), the Gauss curvature of the $\rho\!=\!a$ cylinder in the Kerr space-time {\em diverges} (to plus infinity) at a certain $|z|\!<\!a$ on the $R\!<\!0$ sheet. The explicit formula has the form
\[\frac{^{(2)\!}R}{2}
  =\frac{Ma^2\cos^4\theta\,\sin^2\theta\,\cdot\,({\rm polynomial~of~order~}r^{17})}
        {\Sigma^3{\cal A}^2\,\left[(r^2+a^2)^2+\Delta a^2\right]^2}\]
(where $r$ and $\theta$ are bound by the condition $\rho\!=\!a$, i.e. $\sqrt{r^2\!+\!a^2}\,\sin\theta\!=\!a$),
so the reason of the divergence is the same as in previous section: it occurs where ${\cal A}\!=\!0$, i.e. on the boundary of the chronology-violating region.

\subsection{Kretschmann scalar}

\begin{figure*}
\includegraphics[width=0.9\textwidth]{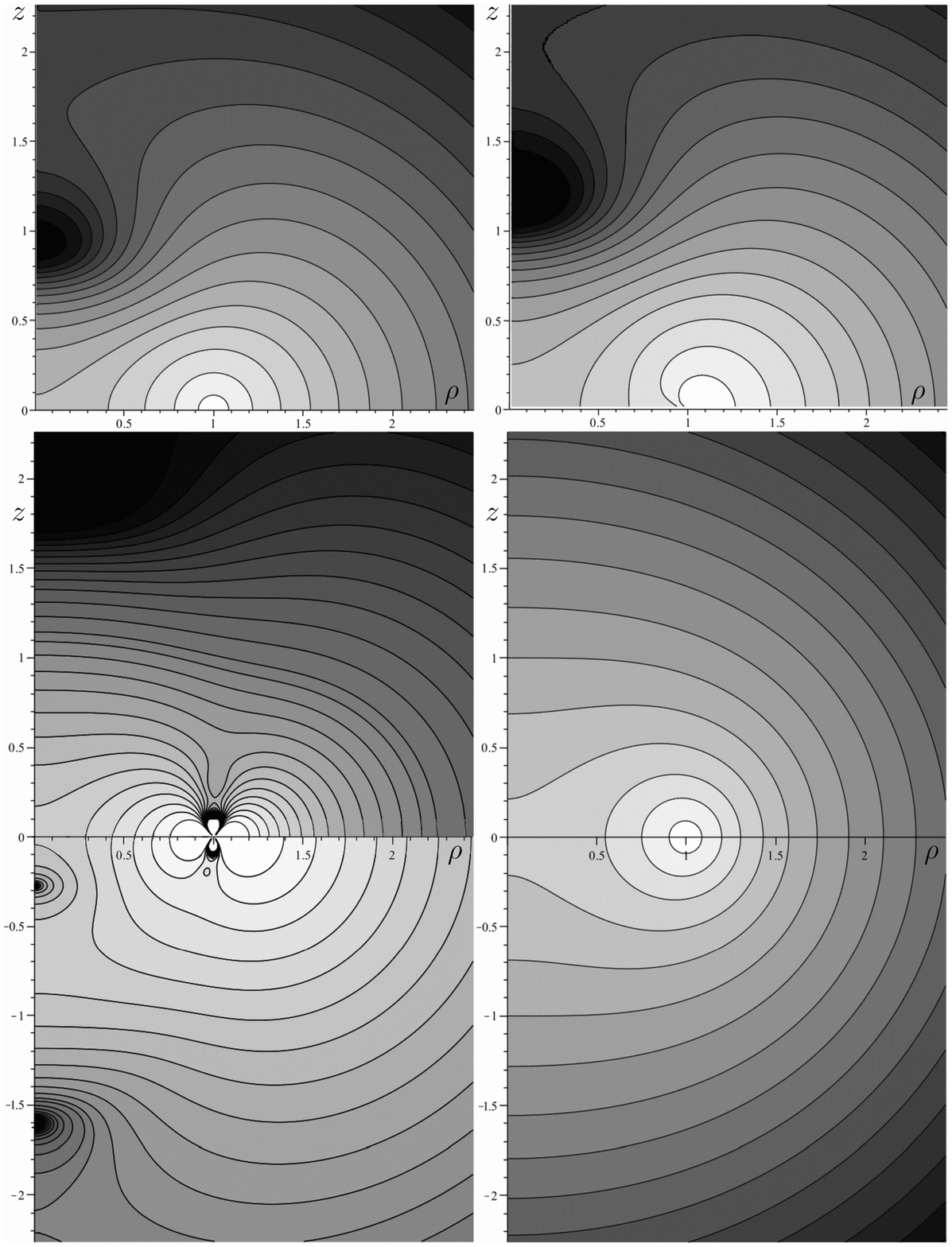}
\caption
{Contours of the Kretschmann scalar within the meridional plane, shown for the Majumdar-Papapetrou ring (top left), the Bach-Weyl ring (top right) and the Appell ring (bottom left), and of the combined Kretschmann/Chern-Pontryagin scalar for the Kerr ring (bottom right). All the rings have $M\!=\!a$. The $(\rho,z)$ axes represent Weyl coordinates for the static rings, while Kerr-Schild coordinates for the Kerr ring. The scalar is everywhere positive, with light/dark shading indicating large/smaller values. In the units of $1/a^4$, the contour levels range from $0.003$ to $100$ for the MP and BW rings, from $0.003$ to $5\cdot 10^6$ for the Appell ring, and from $0.036$ to $8000$ for the Kerr ring.}
\label{Kretschmann}
\end{figure*}

\begin{figure*}
\includegraphics[width=0.9\textwidth]{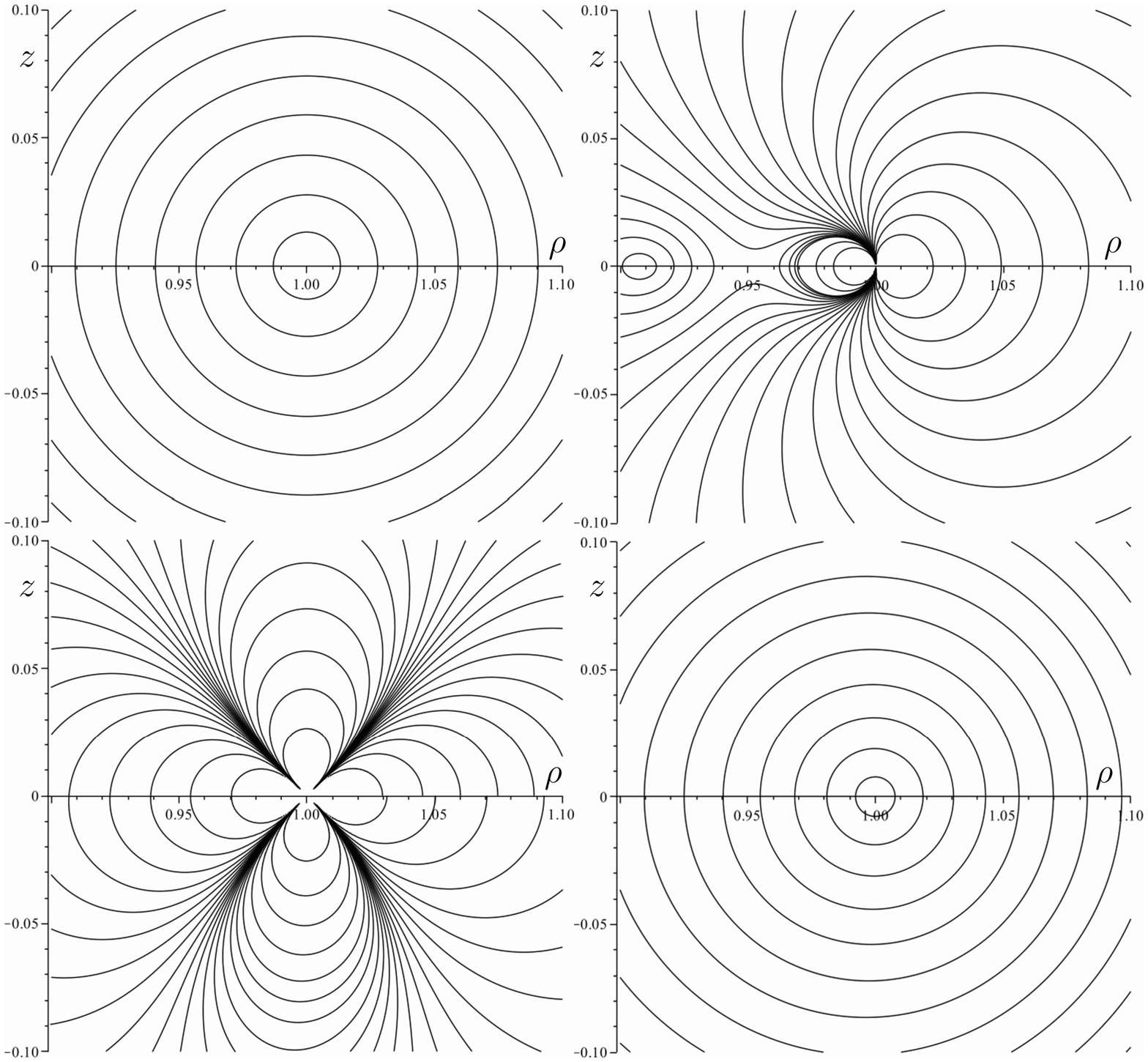}
\caption
{Details of plots shown in figure \ref{Kretschmann}, magnifying the region in a close vicinity of the ring sources ($0.9\!<\!\rho/a\!<\!1.1$, $-0.1\!<\!z/a\!<\!0.1$). Again the Majumdar-Papapetrou ring is at top left, the Bach-Weyl ring at top right, the Appell ring at bottom left and the Kerr ring at bottom right. The rings are clearly very different, with only the Majumdar-Papapetrou and the Kerr rings providing a satisfactory (locally cylindrical) picture. In the units of $1/a^4$, the contour levels now range from $21$ to $5\cdot 10^4$ for the MP ring, from $0.001$ to $10^{15}$ for the BW ring, from $10^{-140}$ to $10^{140}$ for the Appell ring, and from $1880$ to $1.3\cdot 10^7$ for the Kerr ring.}
\label{Kretschmann-detail}
\end{figure*}

Riemann-tensor of the static and axisymmetric space-times has been treated elsewhere (see e.g. \cite{BasovnikS-16}), so we will not repeat it here. What may only be worth summarizing is the special case of the Majumdar-Papapetrou solutions. Due to the relation $\Phi\!=\!\mp N$ between the electrostatic potential and lapse, specific for them, one finds \cite{SemerakB-16}
\begin{equation}
  F_{\mu\nu}F^{\mu\nu}=-2\left[(N_{,\rho})^2+(N_{,z})^2\right]
                      =-2\kappa^2/N^2 \,,
\end{equation}
hence, thanks to the relation between the Ricci tensor and the electromagnetic-field tensor, coming from Einstein equations,
\[R_{\mu\nu}R^{\mu\nu} =(F_{\mu\nu}F^{\mu\nu})^2
                       =\frac{4\kappa^4}{N^4}
                       =4\left[(N_{,\rho})^2+(N_{,z})^2\right]^2 \,.\]
For our MP ring, this is singular at the ring and nowhere else.
Both the electromagnetic invariant and (thus) the Ricci-tensor quadratic scalar are zero at the origin ($\rho\!=\!0$, $z\!=\!0$), like -- naturally -- for the vacuum ring solutions.

The Ricci tensor of the Majumdar-Papapetrou solutions has very simple components,
\begin{align*}
  -R^t_t=R^\phi_\phi &= (N_{,\rho})^2+(N_{,z})^2, \\
  -R^\rho_\rho=R^z_z &= (N_{,\rho})^2-(N_{,z})^2, \\
  R^\rho_z=R^z_\rho  &= -2N_{,\rho}N_{,z} \;, 
\end{align*}
and the non-zero Riemann-tensor components can also be written quite short,
\begin{align*}
  {R^{\rho z}}_{\rho z}={R^{t\phi}}_{t\phi}
     &=(N_{,\rho})^2+(N_{,z})^2-\frac{N}{\rho}\,N_{,\rho} \;, \\
  {R^{t\rho}}_{t\rho}
     &=-(N_{,\rho})^2+(N_{,z})^2-NN_{,\rho\rho} \;, \\
  {R^{tz}}_{tz}
     &=(N_{,\rho})^2-(N_{,z})^2-NN_{,zz} \;, \\
  {R^{\phi\rho}}_{\phi\rho}
     &=(N_{,\rho})^2+(N_{,z})^2-NN_{,zz} \;, \\
  {R^{\phi z}}_{\phi z}
     &=(N_{,\rho})^2+(N_{,z})^2-NN_{,\rho\rho} \;, \\
  {R^{t\rho}}_{tz}
     &=-NN_{,\rho z}-2N_{,\rho}N_{,z} \;, \\
  {R^{\phi\rho}}_{\phi z}
     &=NN_{,\rho z} \;,
\end{align*}
from where the Kretschmann scalar $K\!:=\!R_{\mu\nu\kappa\lambda}R^{\mu\nu\kappa\lambda}$ follows in the form
\begin{align}
  \frac{1}{8}\,K
  =\;& N^2\left[(N_{,\rho\rho})^2+2(N_{,\rho z})^2+(N_{,zz})^2\right]- \nonumber \\
     & -2N\left[(N_{,z})^2 N_{,\rho\rho}+(N_{,\rho})^2 N_{,zz}\right]+ \nonumber \\
     & +4NN_{,\rho}N_{,z}N_{,\rho z}+\frac{N^2}{\rho^2}\,(N_{,\rho})^2+ \nonumber \\
     & +3\left[(N_{,\rho})^2+(N_{,z})^2\right]^2- \nonumber \\
     & -\frac{2N}{\rho}\,N_{,\rho}\left[(N_{,\rho})^2+(N_{,z})^2\right]. \label{Riem^2}
\end{align}

For the MP ring, the result is relatively complicated and even remains so on the $\rho\!=\!0$ axis, but especially at the central point $z\!=\!0$ it simplifies to just
\[K(\rho\!=\!0,z\!=\!0)=\frac{12M^2}{(M+a)^6} \;.\]
This starts from zero for $M\!=\!0$, grows with $M$ up to $64/(243a^4)\doteq 0.2633/a^4$ for $M\!=\!a/2$ and then falls back to zero gradually for larger $M$.
Within the axis the above value represents a maximum, while for the circular interior of the ring it represents a minimum.
The behaviour of the Kretschmann scalar in the meridional plane is illustrated, for $M\!=\!a$, in figures \ref{Kretschmann} (central part of the meridional section) and \ref{Kretschmann-detail} (zoom in on the rings' closest vicinity). Let us add that in the $a\!\rightarrow\!0^+$ limit the above expression gives $12/M^4$ at the very center, which is {\em not} exactly the Kretschmann-scalar value on the extreme Reissner-Nordstr\"om horizon (the latter being $8/M^4$).

Now to compare with the other rings.
For the Bach-Weyl ring, the Kretschmann scalar reduces, on the $\rho\!=\!0$ axis, to
\[K(\rho\!=\!0)=\frac{12M^2\left[(2z^2-a^2)\sqrt{z^2+a^2}-2Mz^2\right]^2}
                     {(z^2+a^2)^6\;\exp\frac{4M}{\sqrt{z^2+a^2}}}\]
which has exactly one zero at certain $|z|\!>\!M$ and the central-point value
\[K(\rho\!=\!0,z\!=\!0)=\frac{12M^2}{a^6\,\exp(4M/a)} \;.\]
In contrast to the MP-ring case, this vanishes for $a\!\rightarrow\!0^+$.
For the Appell ring, one finds
\[K(\rho\!=\!0)=\frac{48M^2\left[(M-R)(R^2-a^2)^2+4a^4 R\right]^2}
                     {(R^2+a^2)^8\;\exp\frac{4MR}{R^2+a^2}} \;;\]
this always vanishes at certain $R\!>\!M$ and also at two other points in the $R\!<\!0$ region (one above and one below $R\!=\!-a$), and it has the central-point value
\[K(\rho\!=\!0,z\!=\!0)=\frac{48M^4}{a^8}\]
which {\em diverges} for $a\!\rightarrow\!0^+$.
Therefore, the $a\!\rightarrow\!0^+$ limits of $K(\rho\!=\!0,z\!=\!0)$ differ considerably for the BW and Appell rings (including the fact that the Appell-ring result is proportional to $M^4/a^8$, whereas those obtained for the other rings -- and also for the Kerr case below -- are proportional to $M^2/a^6$), but this is mainly due to a ``wrong order of limits'': the limit form of the axis result is the same for both and represents {\em correctly} the expression valid for the Curzon solution,
\[\lim\limits_{a\rightarrow 0^+}K(\rho\!=\!0)
  =\frac{48M^2(|z|-M)^2}{z^8\,\exp\frac{4M}{|z|}} \;.\]
(The BW and Appell rings being strongly directional, it is also good to remember that the $a\!\rightarrow\!0^+$ limit of the value at the origin $[\rho\!=\!0,z\!=\!0]$ effectively means approaching the ring from inside.)

For static axisymmetric fields, the Kretschmann scalar and the ``Ricci tensor squared'' are the only independent invariants quadratic in curvature, whereas in the stationary case, one also has the Chern-Pontryagin scalar given by
\[{^*\!}K:={^*\!R}_{\mu\nu\kappa\lambda}R^{\mu\nu\kappa\lambda} \,,\]
representing the ``magnetic'' part of curvature (${^*\!R}_{\mu\nu\kappa\lambda}$ is the Riemann-tensor left dual). Hence, when comparing the static-ring results with the Kerr case, it is appropriate to also take into account this second contribution. This is most reasonably done by considering the modulus of the complex number $(K-{\rm i}\,{^*\!}K)$,
\begin{equation}
  \left|K-{\rm i}\,{^*\!}K\right|=\frac{48M^2}{\Sigma^3} \;,
\end{equation}
which comes out surprisingly simple (much simpler than each of the parts separately); it amounts to $48M^2/a^6$
at the central point [$\theta\!=\!0$,$\,r\!=\!0$], which diverges in the Schwarzschild, $a\!=\!0$ limit.
The comparison is illustrated in figures \ref{Kretschmann} and \ref{Kretschmann-detail} which confirm very calm behaviour of curvature in the Kerr and the MP-ring cases.

Let us again add the $\zeta\!\rightarrow\!\infty$ asymptotics in order to quantify the behaviour at the closest vicinity of the rings:
\begin{align*}
  K &\sim \frac{84\pi^4\,e^{4\zeta}}{M^4\,\zeta^4}
    &\hspace*{-2mm}& \dots\;{\rm MP~ring} \,, \\
    &\sim \frac{16M^6\zeta^6}{\pi^6 a^{10}}\;\exp\left(\frac{8M^2\zeta^2}{\pi^2 a^2}\;e^\zeta\cos\psi\right)
    &\hspace*{-2mm}& \dots\;{\rm BW~ring} \,, \\
    &\sim \frac{M^6}{16384\,a^{10}}\;\exp\left(\frac{M^2}{16a^2}\;e^{2\zeta}\cos 2\psi\right)
    &\hspace*{-2mm}& \dots\;{\rm Appell~ring} \,, \\
  |K&-{\rm i}\,{^*\!}K| \sim \frac{3M^2}{4a^6}\,e^{3\zeta}
    &\hspace*{-2mm}& \dots\;{\rm Kerr~ring} \,;
\end{align*}
it was in fact not necessary to transform the last, combined Kretschmann--Chern-Pontryagin scalar of the Kerr space-time to toroidal coordinates, because the isotropy of $48M^2/\Sigma^3$ is obvious ($\Sigma$ represents square of the ``coordinate distance'' from singularity).

\section{Concluding remarks}
\label{concluding}

Three examples of static and axisymmetric thin-ring sources of general relativity have been analyzed -- the Majumdar-Papapetrou (MP) ring (obtained as a smooth limit of a circular distribution of extremally charged static black holes), the Bach-Weyl (BW) ring and the Appell ring. Despite the ``artificial'' nature of the MP ring, it turns out to generate quite a reasonable field in its vicinity, as opposed to the BW ring (usually referred to as the most ordinary ring source); in particular, the MP ring is not directional (its properties do not much depend on direction from which it is approached), not to mention that its basic dimensions turned out to involve such factors as Catalan's or Ap\'ery's constants -- see (\ref{MP-radius}) and (\ref{MP-area}). Even the rather peculiar, double-sheeted Appell ring (which is also strongly directional) has some characteristics more natural than the BW ring; for example (and in contrast to the BW ring), the circle it encloses has a finite proper radius as well as area, and both go to zero with vanishing Weyl radius.

Besides contrasting the three static rings between each other, we have also compared them with the Kerr ring singularity. The latter is of course {\em rotating} rather than static, and it has a Killing horizon around if $a\!\leq\!M$, but the field it generates has e.g. some similarities with the Appell case (not only that it is double-sheeted too). The Kerr ring turns out to have more satisfactory properties than all the above static rings.

A more general conclusion is that one should be very careful when considering thin (thus singular) sources in general relativity -- much more careful than in the Newtonian theory. This mainly applies if such sources are employed as approximations of an actual distribution of matter in strong-field astrophysical systems. As an example, let us mention the usage of the BW ring in our own study of how the field of a black hole could be deformed by an external source (started in \cite{SemerakZZ-99a} and very recently continued in \cite{BasovnikS-16}). From the properties summarized in table I (of which ``everything diverges'' for the BW ring), it even seems questionable to speak of a source ``surrounded by a ring''. However, it is mainly problematic if the closest vicinity of the ring directly affects a given problem, as for instance when considering the motion of free particles in the ring's field. It has already been pointed out in \cite{SemerakZZ-99b} and then reminded by \cite{D'AfonsecaLO-05}\footnote
{If interested in ring solutions, one should read the Appendix of this paper where several important errors in the literature are admitted. We also added a mistake in \cite{SemerakZZ-99a} and \cite{SemerakS-10}, by giving a wrong expression for the metric function $\lambda$ for the BW ring (it has been corrected in \cite{BasovnikS-16}).}
that near the BW ring the geodesics behave in a non-intuitive manner, and an explanation of such a behaviour was also suggested there. More recently, we studied the geodesic dynamics in the field of an (originally) Schwarzschild black hole surrounded by the BW ring in order to see how strong irregularity (chaotic behaviour) the ring induces, even considering the BW ring (in \cite{SukovaS-13}) as a zero approximation of toroidal configurations of matter known from galactic centers (in particular from that of our Galaxy) and called circumnuclear rings. (See also \cite{WitzanySS-15} where we compared the above relativistic system with its Newtonian counter-part and devoted a special note to the difference between the Newtonian and relativistic version of the ``ordinary'' -- Bach-Weyl -- ring.) We mention these previous results in order to suggest that if employing the BW ring in some (astro)physical problem, it would be more appropriate to at least exclude the closest vicinity of the ring from consideration, to use (e.g.) the MP ring instead,\footnote
{In doing so, one has to remember that the MP-ring space-time is {\em not} a vacuum one, namely that it contains the respective electromagnetic field. However, this does not mean that it cannot be used when studying geodesic motion, for example: one simply has to use particles without electric charge in that case.}
or, in an ideal case, to use {\em extended} (in this case toroidal) sources rather than thin ones -- e.g. those studied in \cite{SachaS-05,VogtL-09,BannikovaVS-11,Fukushima-16}.

When stressing the unsatisfactory, directional nature of the BW ring, probably indicating that a better representation and interpretation could exist, we should also add that a method how to achieve this was suggested by \cite{Taylor-05}. After proving the theorem that in orthogonally transitive, stationary, axisymmetric $C^{4^-}$ (sufficiently smooth) space-times, one can remove, by a coordinate transformation (determined by the level surfaces of Cartan invariants), any directional singularity which is not at an accumulation point of critical points of a scalar curvature invariant with directional behaviour, Taylor demonstrated the applicability of the approach by finding an explicit such transformation for the Curzon and double-Curzon solutions. He also planned in the paper to decipher the structure of the Zipoy-Vorhees, Bach-Weyl and Tomimatsu-Sato space-times, but this has not appeared then (however, see \cite{Taylor-PhD} where some of these other particular cases were tackled, too). In any case, it may be rather nontrivial to reveal a ``true nature'' of a given source (see e.g. \cite{ScottS-86}), and the result of any such transformation may be unsatisfactory in other respects. In other words, the ``true nature'' in general (also) depends on physical setting.

\begin{acknowledgments}
I am grateful for support from the grant GACR-14-10625S of the Czech Science Foundation.
I thank Jan Kub\'{\i}\v{c}ek and Milan Pe\v{s}ta for checking many of the equations (and suggesting simpler forms at several places), Otakar Sv\'{\i}tek and Tayebeh Tahamtan for discussions, and Tom\'a\v{s} Ledvinka for providing an analytical expression of the integral (\ref{area-Appell}) and for a hint concerning the interpretation of the MP ring.
\end{acknowledgments}

\bibliography{rings.bib}

\end{document}